\documentclass[12pt,preprint]{aastex}

\def\Mpch{\ h^{-1}{\rm Mpc}}
\def\kpch{\ h^{-1}{\rm kpc}}
\def\rvir{r_{\rm vir}}
\def\rtwo{r_{200}}

\slugcomment{draft of \today}
\shorttitle{Shock Waves in Cluster Outskirts}
\shortauthors{Hong et al.}

\begin{document}

\title{Shock Waves and Cosmic Ray Acceleration in the Outskirts of Galaxy Clusters}

\author{Sungwook E. Hong$^{1,2}$, Dongsu Ryu$^1$\altaffilmark{,5}, Hyesung Kang$^3$, and Renyue Cen$^4$}
\affil{$^1$Department of Astronomy and Space Science, Chungnam National University, Daejeon 305-764, Korea: swhong@canopus.cnu.ac.kr, ryu@canopus.cnu.ac.kr\\
$^2$Department of Physics, Korea Institute for Advanced Study, Seoul 130-722, Korea
$^3$Department of Earth Sciences, Pusan National University, Busan 609-735, Korea: hskang@pusan.ac.kr\\
$^4$Department of Astrophysical Sciences, Princeton University, NJ 08544: cen@astro.princeton.edu}
\altaffiltext{5}{Corresponding author.}

\begin{abstract}
The outskirts of galaxy clusters are continuously disturbed by mergers and gas infall
along filaments, which in turn induce turbulent flow motions and shock waves.
We examine the properties of shocks that form within $\rtwo$ 
in sample galaxy clusters from structure formation simulations.
While most of these shocks are weak and inefficient accelerators of cosmic rays (CRs), 
there are a number of strong, energetic shocks which can produce large amounts of CR protons
via diffusive shock acceleration.
We show that the energetic shocks reside mostly in the outskirts and a substantial
fraction of them are induced by infall of the warm-hot intergalactic medium from filaments.
As a result, the radial profile of the CR pressure in the intracluster medium
is expected to be broad, dropping off more slowly than that of the gas pressure, 
and might be even temporarily inverted, peaking in the outskirts.
The volume-integrated momentum spectrum of CR protons inside $\rtwo$ has the power-law 
slope of $4.25 - 4.5$, indicating that the average Mach number of the shocks of main CR
production is in the range of $\left< M_s \right>_{\rm CR} \approx 3 - 4$.
We suggest that some radio relics with relatively flat radio spectrum could be explained
by primary electrons accelerated by energetic infall shocks with $M_s \ga 3$ induced in
the cluster outskirts.

\end{abstract}


\maketitle

\section{Introduction}

Clusters of galaxies are the largest gravitationally bound structures that emerged
from hierarchical clustering during the large-scale structure (LSS) formation of
the Universe.
While the central part of many clusters looks relaxed into hydrostatic equilibrium,
especially in X-ray observations \citep[e.g.,][]{mfsv98,vkfj06}, the outskirts around
the virial radius, $\rvir$, are stirred by mergers of substructures and
continuous infall of gas along adjacent filaments
\citep[e.g.,][]{ryu2003,skillman2008,vazza2009a}.
Observational evidence for the deviation from equilibrium in the cluster outskirts can be
seen in the entropy distribution.
The radial profile of $S \equiv kT/n_e^{2/3}$ obtained in X-ray observations follows roughly
$\sim r^{1.1}$ in the inner part of clusters, but beyond it $S$ flattens off and turns down
\citep{vkb05,gfsy09,walker2012,simionescu2013}.
Moreover, according to structure formation simulations,
turbulent flow motions develop during the formation of clusters;
the ratio of turbulence to gas pressure increases outwards and reaches order of 10 \%
in the outskirts of simulated clusters \citep[e.g.,][]{ryu2008,vazza2009b,lau2009}.

Although flow motions are expected to be on average subsonic in the cluster
outskirts\footnote{The ratio of
turbulent to gas pressure of $\sim 30$ \%, for instance, corresponds to the
turbulent Mach number of $\sim 0.42$.},
shock waves have been observed in X-ray as well as in radio.
In X-ray observations, some of sharp discontinuities in the surface brightness are
attributed to shocks, while others are attributed to cold fronts or contact
discontinuities. The physical properties of these shocks including the sonic
Mach number, $M_s$, can be determined using the deprojected temperature and density jumps
\citep{markevitch07}.
Since a shock was found in the so-called bullet cluster (1E 0657–56)
\citep{markevitch02}, about a dozen of shocks have been detected with Chandra, XMM-Newton,
and recently Suzaku \citep[e.g.,][]{russell10,akamatsu12,ob13}.
The shocks identified so far in X-ray observations are mostly weak with
$M_s \sim 1.5 - 3$.

Shocks have been identified also in radio observations through the so-called radio relics
\citep[e.g.,][for reviews]{feretti12,bruggen12}.
Radio relics are the radio structures of megaparsec size observed within the virial radius.
They often show elongated morphologies with sharp edges in one side, and occasionally
come in pairs located in opposite sides of clusters.
Radio emissions from these structures usually exhibit high polarization fractions.
Radio relics are interpreted as shocks, where relativistic electrons emitting synchrotron
radiation are accelerated or re-accelerated.
The properties of radio relic shocks such as $M_s$, magnetic field strength, 
and the age can be estimated from the spectral index and spatial profile of synchrotron emissions
\citep{vanweeren2010,krj2012}.
So far several dozens of radio relics have been observed, and the Mach numbers of
associated shocks are typically in the range of $M_s \sim 1.5 - 4.5$
\citep[e.g.,][]{clarke2006,bonafede2009,vanweeren2010,vanweeren2012}.

In a few cases, shocks were detected both in X-ray and radio observations.
Interestingly, however, the shock characteristics derived from X-ray
observations are not
always consistent with those from radio observations.
For instance, the shock in the so-called sausage relic in CIZA J2242.8+5301
was estimated to have $M_s \simeq 4.6$ in the analysis of radio spectrum 
based on diffusive shock acceleration (DSA) model \citep{vanweeren2010},
but X-ray observations indicated $M_s \simeq 3.2$ \citep{akamatsu13}.
And the shock in the so-called toothbrush relic in 1RXS J0603.3+4214 has
$M_s \simeq 3.3 - 4.6$ according to the radio spectral analysis, but $M_s \la 2$ according
to X-ray observations \citep{vanweeren2012,ogrean13}.
In addition, the positions of shocks identified in radio are often spatially shifted from
those found in X-ray (see the references above).
Resolving these puzzles would require further observations as well as theoretical
understandings of weak collisionless shocks in the intracluster medium (ICM),
which is a high beta plasma with $\beta=P_{\rm th}/P_B\sim 100$ \citep[e.g.,][]{ryu2008}.
Here, $P_{\rm th}$ and $P_B$ are the gas thermal and magnetic pressures,
respectively.

With relatively low Mach numbers as well as elongated morphologies and occasional parings
in opposite sides of clusters, shocks observed in the outskirts are often considered to be
induced by mergers.
The hypothesis of {\it merger shocks} was explored in simulated clusters,
especially for the origin of radio relics
\citep{skillman2011,nuza2012,skillman2013}.
In these studies, shocks in clusters are identified and the injection and acceleration
of cosmic-ray (CR) electrons are modeled.
Then, along with a model for the magnetic field in the intergalactic medium (IGM),
synthetic radio maps are produced and examined.
These studies suggested that merger shocks with sufficient kinetic energy
flux are likely to be responsible for observed radio relics.
However, it was also argued that typical mergers are expected to induce
mostly weak shocks with $M_s\la 3$ and major mergers with similar masses, which are
required to explain, for instance, the sausage relic \citep{vanweeren2010},
tend to generate very weak shocks with $M_s \la 2$ \citep[e.g.,][]{gabici2003}.

The nature and origin of cosmological shocks have been studied extensively,
using numerical simulations for the LSS formation of the Universe
\citep{miniati2000,ryu2003,psej06,kang2007,skillman2008,hoeft08,vazza2009a,bruggen12}.
Shocks are induced as a consequence of hierarchical clustering of nonlinear
structures and can be classified into two categories.
{\it External shocks} form around clusters and filaments of galaxies, when the cool
($T \sim 10^4$ K), tenuous gas in voids accretes onto them.
So the Mach number of external shocks can be very high, reaching up to $M_s \sim 100$ or so.
{\it Internal shocks}, which form inside nonlinear structures, on the other hand,
have lower Mach numbers of $M_s \la 10$ or so, because they form in much hotter
gas that was previously shocked.
It was shown that while a large fraction of internal shocks have $M_s \la 3$, those with
$2 \la M_s \la 4$ are most important in dissipating the shock kinetic energy into heat
in the ICM.
Internal shocks are induced by mergers of substructures, as well as by turbulent flow motions
and by infall of warm gas from filaments to clusters \citep[e.g.,][]{ryu2003}.

{\it Turbulent shocks}, induced by turbulent flow motions, are expect to be weak with
at most $M_s \la 2$, because the root-mean-square (rms) flow motions are subsonic.
{\it Inflall shocks}\footnote{Here we distinguish infall shocks from external accretion
shocks that decelerate never-shocked gas accreting onto clusters and filaments from
void regions. Infall shocks are by nature also accretion shocks that stop previously
shocked gas accreting onto clusters from filaments.},
on the contrary, can have Mach numbers as large as $\sim 10$,
since they form by the infall of the so-called warm-hot intergalactic medium (WHIM)
with $T \approx 10^5 - 10^7$ K to the hot ICM with
$T \approx 10^7 - 10^8$ K.
We note that a continuous infall containing density clumps would be difficult to be
differentiated from a stream of minor mergers with small mass ratios.
But infall shocks clearly differ from merger shocks, which are generated by major mergers,
in the sense that they do not appear as a pair in opposite sides of clusters.
In addition, infall shocks should be found mostly in the cluster outskirts, since the gas
infall from filaments normally stops around the virial radius and does not penetrate into
the core.
So it would be reasonable to conjecture that while weak shocks with $M_s \la 3$ in clusters
are mostly merger or turbulent shocks, stronger shocks with $M_s \ga 3$ found in
the cluster outskirts are likely to be infall shocks.

Shocks that can be categorized as infall shocks were identified in observations
before.
For instance, the radio relic 1253+275 in the Coma cluster was interpreted as an infall
shock formed by a group of galaxies along with the intra-group medium accreting into
the ICM \citep{br11}.
Also the radio structure of NGC 1265 in the Perseus cluster was modeled as the passage
of the galaxy through a shock with $M_s \simeq 4.2$ formed by the infalling WHIM
\citep{pj11}.
However, the properties such as the frequency, spatial distribution, and energetics
of infall shocks
have not been studied in simulations before, partly because the automated
distinction of infall shocks from merger shocks or other types of shocks in simulated
clusters is not trivial.

It is well established that CRs are produced via DSA process at collisionless shocks,
such as interplanetary shocks, supernova remnant shocks, and shocks in clusters
\citep{bell1978,bo78,drury1983}.
Shocks in the LSS of the universe are the primary means through which the
gravitational energy released during the structure formation is dissipated into
the gas entropy, turbulence, magnetic field, and CR particles
\citep[e.g.,][]{ryu2008,ryu2012}.
Post-processing estimations with simulation data for the amount of CR protons produced
in clusters showed that the CR pressure in the ICM may reach up to a few \%
of the gas thermal pressure \citep{ryu2003,kang2007,skillman2008,vazza2009a}.
Observationally, on the other hand, the CR-to-thermal pressure ratio in clusters was
constrained to be less than a few \%, with the upper limits on $\gamma$-ray fluxes
set by Fermi-LAT and VERITAS \citep{acke10,acke13,arlen12}.

In some simulations for the LSS formation, the injection/acceleration of CR protons at
shocks and the spatial advection of the CR pressure were followed self-consistently in
run-time \citep[e.g.,][]{miniati2001,pfrommer2007,vazza2012}.
\citet{pfrommer2007} and \citet{vazza2012}, adopting specific DSA efficiency models,
showed that the CR acceleration occurs mostly in the cluster outskirts.
Because of long lifetime and slow particle diffusion \citep[e.g.,][]{bbp97}, the
CR protons accelerated in the outskirts are likely to be
contained in clusters and accumulated
in the ICM over cosmological time scales.
But they can be advected with turbulent flows toward the central part
of the cluster \citep[see, e.g.,][]{epms11}.
For simplicity, 
let us assume that the transport of CR protons due to flow motions can be
approximated by turbulent diffusion, then
it could be be described by 
$\partial Q({\vec r},t)/\partial t =  {\vec\nabla}\cdot[D({\vec r},t){\vec\nabla}Q({\vec r},t)]$,
where $Q({\vec r},t)$ is the density of CR protons and $D({\vec r},t)$ is the turbulent
diffusion coefficient.
If only the radial diffusion is considered and the diffusion coefficient is approximated
as $D(r,t) \sim r V(r)$, where $V(r)$ is the average flow speed at $r$, 
then the advection time scale can be estimated rather roughly as
$\tau_{\rm adv}  \sim r^2 / D \sim r / V(r)$.
In the cluster outskirts, typically 
$r \sim 1 \Mpch$ and $V(r) \sim$ a few $\times\ 100$ km/s, 
so $\tau_{\rm adv} \sim$ a few $\times\ 10^9$ yrs.
This is a substantial fraction of the age of the universe, implying that it would take
a while for CR protons produced at energetic shocks in the outskirts to reach the core region.
As a result, the radial distribution of the CR pressure would be broad, dropping
off more slowly than that of the gas thermal pressure in the outskirts.
\citet{vazza2012} also showed that the CR pressure distribution could be temporarily
inverted, that is, the CR pressure can increase outwards.

\citet{brunetti2012}, on the other hand, 
attempted to constrain the radial distributions of nonthermal components
(including the CR proton energy density) in the Coma cluster
by combining radio observations 
with recent Fermi-LAT $\gamma$-ray observations and with Faraday rotation measure (RM) data.
They argued that the model based on the turbulent acceleration of secondary electrons would
best reproduce the radio halo of the Coma cluster with the CR energy density that scales
with the thermal energy density as $\varepsilon_{\rm CR} \propto \varepsilon_{\rm th}^{\theta}$
with $\theta\approx -0.1$ to $-0.35$, implying that $\varepsilon_{\rm CR}$ is higher at lower
$\varepsilon_{\rm th}$.
The outer region of the Coma cluster is strongly disturbed by ongoing mergers and
infalls \citep[e.g.,][]{simionescu2013}, so it would be probably one of rare cases with
this kind of inversion of the $\varepsilon_{\rm CR}$ profile.
But these indicate that the partitioning of thermal and CR energies (and possibly
turbulent and magnetic field energies too) could be very different in different parts of
clusters.

In this paper, we study shocks within the virial radius in a sample of clusters taken
from LSS formation simulations.
Specifically, we examine the properties of {\it energetic shocks} with relatively high
Mach number and high shock kinetic energy flux that can produce large amounts of CR protons
via DSA.
The plan of this paper is as follows.
In Section 2, numerical details are presented.
In Section 3, the properties and nature of shocks in the cluster outskirts are described.
In Section 4, the properties of CRs produced at energetic shocks are described.
Discussion is given in Section 5, and summary follows in Section 5.

\section{Numerical Details}

\subsection{Cluster Sample}

To produce a sample of galaxy clusters, we performed simulations of the LSS formation,
using a Particle-Mesh/Eulerian cosmological hydrodynamics code described in \citet{ryu1993}.
A standard $\Lambda$CDM cosmological model was assumed with the following parameters:
baryon density $\Omega_\mathrm{BM} = 0.044$, dark matter density $\Omega_\mathrm{DM} = 0.236$, 
cosmological constant $\Omega_\Lambda = 0.72$, Hubble parameter
$h \equiv H_0 / (100 \mathrm{km/s/Mpc}) = 0.7$, rms density fluctuation $\sigma_8 = 0.82$,
and primordial spectral index $n = 0.96$.
These parameters are consistent with the WMAP7 data \citep{komatsu2011}.
Cubic boxes of comoving sizes of $100$ and $200 \Mpch$ with periodic boundaries were
employed and divided into $1024^3$ grid zones with uniform spatial resolutions of
$\Delta l = 97.7$ and $195.3 \kpch$, respectively.
Nongravitational processes such as radiative cooling, star formation and feedback, and
reionization of the IGM were not considered.
Instead, a temperature floor was set to be $T_{\rm min}=10^4$K for the gas in voids,
assuming that the unshocked gas outside nonlinear structures was uniformly heated by
reionization.
To compensate the cosmic variance and acquire an enough number of clusters, 16 runs
with different realizations of initial condition were performed for each of $100$ and
$200 \Mpch$ boxes (so the total number of runs is 32).

In addition, we used a higher resolution simulation with $2048^3$ grid zones in box
of $100 \Mpch$ comoving size ($\Delta l = 48.8 \kpch$), to mainly examine the
resolution effects.
This simulation was performed with a numerical code described in \citet{llc08}, 
adopting the same set of cosmological parameters except $\Omega_\mathrm{BM} = 0.046$.
It is basically the same simulation reported in \citet{cen2011}, but with the box size
of $100 \Mpch$ instead of $50 \Mpch$.
The simulation includes a mild feedback from star formation (low galactic superwind feedback
of \citet{cen2011}) and cooling/heating processes.
\citet{kang2007} examined the effects of a similar feedback and radiative processes on
the properties of shocks in the LSS.
They showed that the dynamics and energetics of shocks are governed primarily
by the gravity of matters, so mild feedback and cooling do not significantly affect
the statistics of the shocks in the ICM \citep[see][for the case that the feedback is
stronger and its effects are more important]{pfrommer2007}.

In the simulation data, we identified clusters as the volumes with high X-ray luminosity
\citep[see][for details]{kang1994}.
For each identified cluster, we calculated the gas mass, $M_{\rm cl}$, and the X-ray
emission-weighted average temperature, $T_\mathrm{X}$, inside $\rtwo$ from the cluster
center that locates at the peak of X-ray emissivity.
Here $\rtwo$ is defined as the radius within which the gas overdensity is 200 times the
{\it mean gas density} (not the critical density) of the universe.\footnote{The relation
between $\rtwo$ and $\rvir$ is rather complicated and depends on cosmological parameters
\citep[e.g.,][]{ns97,bn98,enf98}.
For the parameters we employed, approximately $\rtwo \simeq 1.3\ \rvir$.}
We built our sample with clusters of $T_\mathrm{X} \geq 2$ keV from $100 \Mpch$
box simulations and those of $T_\mathrm{X} \geq 4$ keV from $200 \Mpch$ box simulations,
by optimizing the resolution limitation and the size of cluster sample;
125 clusters were identified from 16 simulations of $100 \Mpch$ box with $1024^3$ zones,
94 clusters from 16 simulations of $200 \Mpch$ box with $1024^3$ zones, and 9 clusters
from one simulation of $100 \Mpch$ box with $2048^3$ zones.
Figure 1 shows the radius-mass relation and the radius-temperature relation 
of the total 228 clusters in our sample.
The simulated clusters have $\rtwo\approx 1 - 3 \Mpch$,
$M_{\rm cl}\approx 10^{13} - 10^{14} M_{\odot}$, and $T_\mathrm{X} \approx 2 - 10$ keV.
From the virial theorem, the mass, temperature, and radius of relaxed clusters are expected
to follow $M_{\rm vir} \propto \rvir^3$ and $T_{\rm vir} \propto M_{\rm vir} / \rvir$
or $\rvir \propto M_{\rm vir}^{1/3} \propto T_{\rm vir}^{1/2}$ \citep[e.g.,][]{peeb80}.
As can be seen from Figure 1, overall $M_{\rm cl}$, $T_\mathrm{X}$, and $\rtwo$ of our
clusters follow these relations, but there are scatters, a part of which are
caused by dynamical activities in the cluster outskirts.

We point that with uniform grids of $\Delta l = 48.8 - 195.3 \kpch$, our simulated
clusters have poorer resolution than those generated using adaptive mesh refinement
(AMR) or SPH codes (see Introduction for references), especially in the core regions.
This means that shocks in the inner regions of clusters may not be fully reproduced.
But those shocks in high density regions are expected to be weak with low Mach numbers of
$M_s \la 2 - 3$ \citep[e.g.,][]{vazza2011}, so their contribution to the production
of CRs would not be significant (see the discussion in Section 2.3).
Since this paper focuses on relatively strong, CR-producing shocks in the outskirts,
having an uniform resolution throughout the entire simulation volume
should be actually an advantage.

\subsection{Shock Identification}

A number of algorithms that can be applied to the identification of shocks in
structure formation simulation data
have been suggested \citep[e.g.,][]{ryu2003,psej06,skillman2008,vazza2009a}.
They all employed the Rankine-Hugoniot jump conditions but in slightly different ways.
Although there are some differences, the properties of identified shocks in the LSS
by different algorithms are overall consistent with each other
\citep[e.g.,][for a comparison study]{vazza2011}.
Here we adopted the algorithm suggested by \citet{ryu2003}.

A series of the following one-dimensional procedures are first gone through for three
primary directions.
The grid zones are tagged as `shocked' if they fulfill all of the following conditions:
(1) $\vec{\nabla}\cdot\vec{v} < 0$, i.e., the local flow is converging,
(2) $|\Delta \mathrm{log} T| > 0.11$, i.e., the Mach number is greater than 1.3,
and (3) $\Delta T \times \Delta \rho > 0$, i.e., the gradients of temperature and density
have the same sign.
The central difference is defined as $\Delta Q_i \equiv Q_{i+1} - Q_{i-1}$ for the quantity 
$Q_i$ in the zone $i$.
A shock in simulation usually spreads over several grid zones, and the `shock center' is
defined as the grid zone with minimum $\vec{\nabla}\cdot\vec{v}$.
While \citet{ryu2003} used the condition $\Delta T \times \Delta s > 0$ (where $s$ is
the entropy), here we used the condition $\Delta T \times \Delta \rho > 0$ in order
to exclude the possible misidentification of contact discontinuities.
We found that the current method with the condition $\Delta T \times \Delta \rho > 0$
may miss some of weak shocks, but
the overall statistics of identified shocks are not significantly affected.
The Mach number at the shock center, $M_s$, is calculated by solving the relation for
the gas temperature jump along the three primary directions:
${T_2}/{T_1} = (5M_s^2 -1)(M_s^2 + 3)/(16 M_s^2)$.
Hereafter, the subscripts 1 and 2 indicate the preshock and postshock quantities, respectively.
Then the Mach number of the shock center is assigned as the maximum
value of the three Mach numbers, i.e., $M_s={\rm max} (M_{s,x}, M_{s,y}, M_{s,z})$.
Because of complex flow patterns and shock surface topologies, very weak shocks are
difficult to be identified unequivocally, so only shocks with $M_s \ge 1.5$
are considered.
Hereafter, we refer to a grid zone with assigned $M_s$  as ``a shock'', which
represent a small patch with an area of $(\Delta l)^2$.
A shock surface normally consists of many of these shocks (or shock zones),
so the total number of identified shocks multiplied by $(\Delta l)^2$ is effectively
equal to the total area of shock surfaces contained in a given volume.

Once the Mach number is determined, the shock speed and the shock kinetic flux are
estimated as $v_1=M_s (\gamma P_{\rm th,1}/\rho_1)^{1/2}$ and
$f_{\rm kin}=(1/2)\rho_1 v_1^3$, where $\gamma=5/3$ is the gas adiabatic index.

\subsection{Energy Dissipation at Shocks}

If no CRs are produced at a shock, the gas thermalization efficiency can be calculated 
directly from the Rankine-Hugoniot relation as
$\delta_0 (M_s) = {[e_\mathrm{th,2}-e_\mathrm{th,1} (\rho_2 / \rho_1 )^{\gamma} ] v_2}/
f_{\rm kin}$,
where $e_\mathrm{th}$ is the internal energy density.
Note that the second term inside the brackets subtracts the effect of adiabatic 
compression that occurs at a shock as well as the contribution of the thermal energy
flux entering the shock.
Then the generation of heat can be estimated with the thermal energy flux, 
$f_\mathrm{th} = \delta_0(M_s) \times f_{\rm kin}$.
However, if CRs are accelerated via DSA, a fraction of the shock kinetic energy is
transferred to the CR component and the resulting
thermalization efficiency is reduced, i.e., $\delta(M_s) < \delta_0 (M_s)$.
With $\eta(M_s)$ defined as the CR-acceleration efficiency \citep[e.g.,][]{ryu2003,kj07},
the acceleration of CR protons at shocks can be quantified with the CR energy flux,
$f_\mathrm{CR} = \eta(M_s) \times f_{\rm kin}$.

At the moment it is not possible to predict $\delta(M_s)$ and $\eta(M_s)$ from first
principles, because complex wave-particle plasma interactions governing 
the CR injection and acceleration at collisonless shocks are not fully understood.
It has been recognized that the magnetic field amplification (MFA) due to CR streaming
instabilities and the Alfv\'enic drift of scattering centers in the amplified field
play significant roles in DSA at astrophysical shocks such as supernova remnant shocks
\citep[e.g.,][]{lucek00,bell04,schure2012,caprioli2012,kang2012}.
Through numerical simulations of nonlinear DSA for shocks expected in the LSS,
\citet{kang2013} has shown that if self-amplification of magnetic fields and fast
Alfv\'enic drift in the shock precursor are implemented into the standard DSA theory,
the CR energy spectrum is steepened and the CR-acceleration efficiency is reduced,
compared to the cases without including those processes.
Here, we adopted $\delta(M_s)$ and $\eta(M_s)$ of \citet{kang2013}.

Figure 2 shows the curves that fit the values of $\delta(M_s)$ and $\eta(M_s)$ for shocks 
that form in a weakly magnetized plasma with $\beta = 100$
and $n_{H,0}=10^{-4}\ {\rm cm^{-3}}$.
The high value of $\beta \sim 100$ is expected for plasmas in the ICM, as noted in
Introduction.
Both $\delta(M_s)$ and $\eta(M_s)$ increase as $M_s$ increases, and asymptote to $0.45$
and $0.22$, respectively, for strong shocks with $M_s \gtrsim 10$.
Compared to the previous estimate of $\eta \approx 0.55$ at strong shocks, given in
\citet{kang2007} where MFA and Alfvenic drift were not considered, the newly estimated
CR-acceleration efficiency is smaller by a factor of $\sim 2.5$ for $M_s \ga 10$.
Also the CR-acceleration is inefficient at weak shocks with $M_s \la 3$
in our new estimation.

The steepening of CR spectrum due to Alfv\'enic drift and the ensuing reduction
of $\eta(M_s)$ become important, only if the magnetic field is strong enough so that the
Alfv\'enic speed is substantial, that is, $V_A \ga 0.1\ v_1$ ($M_A \la 10$),
where  $V_A$ is the Alfv\'enic speed in the amplified magnetic field in the shock precursor
\citep{caprioli2012,kang2012}.
At weak shocks ($M_s \la 3$), however, 
MFA would be inefficient, so $V_A \approx V_{A,0} = \sqrt{2/(\beta \gamma)}\ c_s$,
where  $V_{A,0}$ is the Alfv\'enic speed in the background magnetic field.
For $\beta = 100$, then $M_{A} \approx M_{A,0} = v_1/V_{A,0} \approx 10\ M_s$, 
and so the Alfv\'enic drift effect would be only mildly important at weak shocks.
At strong shocks, on the other hand, the diffusive CR pressure induces a precursor, 
in which the upstream flow is decelerated and adiabatically compressed,
and the streaming CRs amplify significantly the turbulent magnetic fields
\citep{bell04}.
According to the MFA prescription adopted in \citet{kang2013},
the MFA factor increases with $M_{A,0}$ and can be approximated as
$B_1/B_0\approx 0.1\ M_{A,0}$,
where $B_0$ and $B_1$ are the magnetic field strengths in the background medium 
and immediately upstream of the shock, respectively.
Then, the Alfv\'enic Mach number defined by the amplified magnetic field becomes
$M_{A,1}=M_{A,0}(B_0/B_1) \approx 10$ (independent of the plasma $\beta$),
so the Alfv\'enic drift is expected to be important at strong CR modified shocks.
We note, however, that relevant plasma physical processes, such as the injection of CRs
as well as MFA and Alfv\'enic drift, are not well understood, 
so any attempts to predict
the DSA efficiency involve large uncertainties, especially for weak shocks. 
So the dissipation efficiencies given in Figure 2 should be
taken as rough estimates.

In this paper, we do not consider the re-acceleration of pre-existing CRs.
We note, however, that it could be important especially for weak shocks with
$M_s \la 3$ \citep{kang2011,kang2013}.
\citet{krj2012} and \citet{pop13}, for instance, argued that the re-acceleration of
pre-existing CR electrons would be operating at radio relics associated with weak 
structure formation shocks. 

\section{Properties and Nature of Shocks in Cluster Outskirts}

We first examine the Mach number and energetics of shocks within and around the virial
radius, specifically in $r \leq \rtwo \approx 1.3\ \rvir$, of simulated clusters.
Figure 3 shows the frequency distribution of shocks (i.e., zones with assigned $M_s$)
as a function of Mach number and CR energy flux for the shocks found
in 134 clusters from
$100 \Mpch$ box simulations with $1024^3$ and $2048^3$ grid zones.
The energetics of shocks is quantified with the CR energy flux, $f_{\rm CR}$.
Most of these shocks are internal shocks which are produced in the hot ICM of clusters
or in the WHIM of filaments, according to the classification of \citet{ryu2003}.
As previously shown, they are mostly weak with $M_s \la 3$.
The fractions of shocks with relatively high Mach numbers of $M_s \geq 3$, $\geq 4$,
and $\geq 5$ are $\sim 19 \%$, $\sim 8 \%$, and $\sim 4.5 \%$, respectively,
among all the shocks with $M_s \ge 1.5$ in $100$ and $200 \Mpch$ box simulations.
We categorize shocks or shock zones with
$f_\mathrm{CR} \geq 10^{42}\,\mathrm{ergs}\ \mathrm{s}^{-1} (h^{-1}{\rm Mpc})^{-2}$
as {\it energetic shocks}.
We note that the shocks responsible for observed radio relics are estimated to have
the total kinetic energy flux of $\sim 10^{44} - 10^{45}\,\mathrm{ergs}\ \mathrm{s}^{-1}$
over the entire shock surface of $\sim (h^{-1}{\rm Mpc})^2$ or so
\citep[e.g.,][]{vanweeren2010,krj2012}.
Considering $\eta \approx 10^{-2}$ at $M_s \sim 3$, shocks with
$f_\mathrm{CR} \ga 10^{42}\,\mathrm{ergs}\ \mathrm{s}^{-1} (h^{-1}{\rm Mpc})^{-2}$
can be considered to be energetic enough to be parts of observable radio relics.
The fraction of energetic shocks is $\sim 21 \%$ among all the shocks.
And the fraction of shocks with $M_s \geq 3$ and
$f_\mathrm{CR} \ge 10^{42}\,\mathrm{ergs}\ \mathrm{s}^{-1} (h^{-1}{\rm Mpc})^{-2}$
is $\sim 16 \%$.

Shocks with lower $M_s$ form on average close to the core with higher
gas density (see Figure 4 below) and so have larger $f_{\rm kin}$, but they have lower $\eta$.
Stronger shocks, on the other hand, form mostly in the outskirts and have lower $f_{\rm kin}$,
but they have higher $\eta$.
Such tendencies are reflected in the relation between $f_{\rm CR}=\eta \cdot f_{\rm kin}$
and $M_s$ in Figure 3.
For weak shocks with $M_s \la 3$, the CR acceleration efficiency
increases steeply with $M_s$, while the shock kinetic
energy flux varies only mildly.  So $f_{\rm CR}$ increases strongly 
with $M_s$, resulting in a relatively robust correlation  between the two quantities.
For shocks with $M_s \ga 3$, the dependence of $\eta$ on $M_s$ becomes
much softer, while the variance of $f_{\rm kin}$ increases.
So the correlation between $f_{\rm CR}$ and $M_s$ substantially weakens.
We find that shocks with the largest $f_{\rm CR}$ have 
typically $M_s \approx 3 - 5$, which
interestingly coincides with the Mach numbers of strong radio relic shocks
(see Introduction).

Figure 4 shows two-dimensional slices of three sample clusters with the X-ray
emission-weighted temperatures of $T_{\rm X} =2.7$ keV (left), $2.5$ keV (middle) and
$2.4$ keV (right), respectively, at present ($z=0$).
The slices were chosen to highlight the shock structures, so they pass through short
comoving distances of $0.24-0.28 \Mpch$ from the cluster centers.
The CR luminosity, shown in the bottom panels, is $F_{\rm CR} = f_{\rm CR} (\Delta l)^2$
at the comoving surfaces of shocks.
Hereafter the shock with the largest $f_\mathrm{CR}$ among shocks in each cluster will be
called the {\it most energetic shock} (MES).
Thick arrows in the $M_s$ and $F_{\rm CR}$ panels of Figure 4 point the MESs of each cluster. 
The MESs are located at $r \approx 1.1 \Mpch$ (left), $1.5 \Mpch$ (middle) and $1.1 \Mpch$
(right) from the cluster centers, which correspond to $0.60\ \rtwo$ (left), $0.94\ \rtwo$
(middle) and $0.63\ \rtwo$ (right), respectively.
The spatial distribution of $M_s$ tells that strong shocks are found in the outer regions
of clusters.
Obviously, the strongest shocks are external shocks 
that form in the accreting gas from voids ($T\approx 10^4$K) \citep[see][]{ryu2003}.
But as noted in Introduction, owing to low density, they are energetically unimportant,
so we are not concerned about those external shock in this paper.
Energetic shocks that produce large amounts of CRs are internal shocks and they
reside mostly in the outskirts of the clusters, as shown in the distribution of $F_{\rm CR}$.

The distributions of $\rho$, $\vec{v}$, $T$, and $M_s$ indicate that the structures including
the MESs in Figure 4 look like infall shocks that form by the infall of gas from filaments.
Those infall shocks are energetic enough to penetrate into the region inside the virial
radius where the gas density is relatively high, indicating that their shock kinetic energy
flux is large.
They are also relatively strong with $M_s \sim 5-7$, so they are efficient CR accelerators.
These characteristics make the infall shocks the MESs in the clusters shown here.
We point that not all filaments induce infall shocks inside the virialized regions of
clusters.
Also the cross sectional areas of penetrated filaments are small, compared to the surface
area of virialized regions $\sim 4\pi \rvir^2$.
So the energetic infall shocks inside the virial radius should account for a small fraction
of internal shocks in the ICM.
The distribution of $F_{\rm CR}$ in Figure 4, however, indicates that infall shocks
in the outskirts could be responsible for a significant fraction of CR production in the
clusters (see below).

Figure 5 shows the time evolution of the cluster shown in the right column of Figure 4, 
demonstrating how the cluster has evolved dynamically and how various types of shocks have
been induced. Strong external shocks persist at the comoving distance $3 - 5\Mpch$ away from
the cluster center,
while numerous internal shocks form and disappear in a dynamical timescale of  $\sim 1$ Gyr.
One can see that the cluster experienced a merger at the look-back time of $2 \times 10^9$
years, producing several merger shocks.
Then infall from an attached filament in the south-west direction followed, and penetrated
into the core region.
It was halted by an energetic infall shock that developed around the look-back time of 1 Gyr
and lasted to the present epoch.

To quantify the statistics of infall shocks, we separated them from other shocks.
Infall shocks are defined as those that decelerate the WHIM accreting from filaments to a cluster
(see Introduction).
So we employed the following criteria for infall shocks in $r \leq \rtwo$ using the entropy and
density of the preshock gas and the sonic Mach number:
(1) $\log[T_1(\rm K)/(\rho_1/\left<\rho\right>)^{2/3}] \le 5.3$,
(2) $\rho_1/\left<\rho\right> \le 10^3$, and (3) $M_s \ge 3$.
Note that the first criterion is an entropy condition.
Figure 6 shows the volumetric distribution of the gas from $100 \Mpch$ box
simulations with $1024^3$ grid zones.
The domain demarcated by the first and second criteria 
does not coincide with the conventional definition of the WHIM, $10^5$ K $\le T \le 10^7$ K.
But visual inspections indicated that the above three criteria pick up infall shocks in
the region of $r \le r_{200}$ best among different criteria we have tried.
Figure 7 shows a slice displaying the positions of shocks identified as infall shocks according to
the above criteria as well as those which are not infall shocks, for the cluster shown in the right
column of Figure 4.
We note that the morphology of shock surfaces could be quite complicated, depending on the dynamical
history of clusters.
In the cluster shown, for instance, the infall shock including the MES is connected
to a larger shock surface, a portion of which is a (not-infall) shock expanding from the core to
the outskirt.
In general a connected shock surface can consist of a number of infall and not-infall shocks.

With the above criteria, we found that, among all the shocks with $M_s \ge 1.5$ 
located in $r \le r_{200}$ of the sample clusters, about $10 \%$ are classified as infall shocks.
So most of ICM shocks are merger shocks or turbulent shocks (see Introduction).
As noted in Figure 3, $\sim 19 \%$ of the  identified shocks  
have $M_s \ge 3$, so about a half ($\sim 55 \%$) of them are infall shocks.
Identifying merger shocks would require the examination of the time evolution
of cluster dynamics, which we did not attempt here.

Among the energetic shocks with
$f_\mathrm{CR} \geq 10^{42}\,\mathrm{ergs}\ \mathrm{s}^{-1} (h^{-1}{\rm Mpc})^{-2}$,
the fraction of infall shocks is $\sim 44 \%$.
And among the MESs in 228 sample clusters, 177 are infall shocks;
i.e., $\sim 78 \%$ of the MESs are infall shocks in our sample.
We expect the MESs that are not infall shocks to be merger shocks.
The top panels of Figure 8 show the distributions of the radial position, 
$r_\mathrm{MES}$, and the Mach number, $M_\mathrm{MES}$, of the MESs.
The MESs that are infall shocks (red histogram, MES-ISs hereafter) are distributed over all
radius, peaking at $r_\mathrm{MES}/\rtwo \sim 0.5$.
On the other hand, the MESs that are not infall shocks (blue histogram, MES-NISs hereafter)
are mostly found at inner parts of clusters.
For the MES-ISs, the Mach number distribution peaks around $4 \la M_\mathrm{MES} \la 5$, 
and decreases sharply for smaller $M_\mathrm{MES}$ but extends to larger $M_\mathrm{MES}$.
For the MES-NISs, the Mach number distribution is mostly limited to $M_\mathrm{MES} \la 5$.
The figure demonstrates that the MES-ISs are found mainly at outer parts of clusters and they have
higher Mach numbers than the MES-NISs, as expected.
We attempt to find correlations among $r_\mathrm{MES}$, $T_\mathrm{X}$, and
$M_\mathrm{MES}$ in the bottom panels of Figure 8.
It appears that there is no noticeable correlation between $r_\mathrm{MES}$ and
$T_\mathrm{X}$.
But $r_\mathrm{MES}$ tends to be larger at larger $M_\mathrm{MES}$, confirming that
higher Mach number shocks form at outer regions of clusters.

\section{Cosmic Ray Production at Shocks in Cluster Outskirts}

We next examine the spatial characteristics of the CR proton production in clusters.
Figure 9 shows the radial distributions of the maximum Mach number, $M_{s,{\rm max}}$,
the CR luminosity per unit radius, $L_{\rm CR}$, and the fraction of CR luminosity due to
infall shocks, $L_{\rm CR,infall}/L_{\rm CR}$, in four sample clusters.
Here, $M_{s,{\rm max}}$ is defined as the highest Mach number of shocks located in the
bin of $[r, r+dr]$, while $L_{\rm CR}$ is the sum of $F_{\rm CR}$ of shocks in the bin
divided by the width $dr$.
Note that $L_{\rm CR}$
represents the amount of CR energy produced per unit time per unit length and has a rather
unusual unit, ${\rm ergs\ s^{-1}}(\Mpch)^{-1}$ \footnote{
The volume-averaged CR energy production rate per unit volume at given radius is
$L_{\rm CR}/4\pi r^2$.
Due to the large range of $L_{\rm CR}$, the radial distribution of $L_{\rm CR}/4\pi r^2$ looks
similar to that of $L_{\rm CR}$.}.
The vertical dashed lines mark the radial bins that contain the MESs.

The distribution of $M_{s,{\rm max}}$ demonstrates that on average shocks tend to be stronger
in the outer regions of clusters, as expected.
The CR luminosity $L_{\rm CR}$ is dominated by energetic shocks in a given radial bin,
and energetic shocks are found mostly in the outskirts, so $L_{\rm CR}$ is higher there.
The MESs and the peaks of $L_{\rm CR}$ are located in $r \ga 0.4\ \rtwo$ 
(although the two do not necessarily coincide), indicating
that CRs are produced mostly at the outer regions of clusters.
These findings are consistent with the previous studies using the structure formation simulations
in which the production of CR protons was explicitly followed in runtime
\citep{pfrommer2007,vazza2012}.
With high CR production at the outskirts, we expect the the radial profile of the CR pressure
is flatter than that of the gas pressure and could be even inverted \citep[see][]{brunetti2012},
as mentioned in the Introduction.

The distribution of $L_{\rm CR, infall}/L_{\rm CR}$ shows that infall shocks contribute to
the CR production by a large fraction, especially in the outskirts.
As a matter of fact, we estimate that infall shocks produce $\sim 68 \%$
of CRs in $r \leq \rtwo$, when summed for all 228 clusters in our sample.
Recall that the fraction of infall shocks among those with $M_s \ge 3$ inside the volume of 
$r \leq \rtwo$ is $\sim 55 \%$.

We also examine the momentum distribution of the CR protons, 
which are expected to be produced in the outskirts
and then mixed in the ICM via turbulent flow motions (see Introduction).
For each shock with $M_s$, we adopted the test-particle power-law distribution function,
$f_p(p) \propto p^{-q}$,
where $q = 4M_s^2 / (M_s^2-1)$ \citep{drury1983}.
Note that the CR acceleration at cluster shocks with $M_s \la 5$ is reasonably
described by the test-particle solution \citep{kang2010}.
In our sample, $\sim 96 \%$ of shocks have $M_s \leq 5$ and shocks with the largest $f_{\rm CR}$
have typically $M_s \approx 3 - 5$ (see Section 3).
So the test-particle solution should give reasonable results.
For each sample cluster,
the volume-integrated, momentum distribution function of CR protons within the
radius $r$, $f_p(p,<r)$, is estimated as follows:
(1) At each shock zone $i$, the power-law function,
$f_{p,i}(p) = f_0 p^{-q}$, is normalized with the CR energy flux at the shock as
$f_{{\rm CR},i} = 4\pi m_p c^2 \cdot v_2 \int_{p_{\rm min}}^{p_{\rm max}} (\sqrt{p^2+1} - 1) f_{p,i}(p) d^3p$.
Here, $p$ is expressed in unit of $m_p c$, and $p_{\rm min} = 10^{-2} m_p c$ and $p_{\rm max} = \infty$
are assumed.
2) The volume-integrated momentum distribution function is calculated by adding up $f_{p,i}$ 
for all shocks inside $r$, i.e.,
$f_p(p, <r) = \sum_{r_i < r} f_{p,i}(p)$.
3) Then, the slope of the integrated momentum distribution function is estimated
by fitting $f(p, <r)$ to a power-law form with the slope ${\bar q}(<r)$.

The left panel of Figure 10 shows the values of $\bar{q}(<r)$ calculated for some of sample
clusters\footnote{
The profile of $\bar{q}(r)$, the slope of the momentum distribution of CR protons produced at
shocks in the bin of $[r,r+dr]$, is similar to that of $\bar{q}(<r)$, since the CR production
is dominated by shocks in the outer regions of clusters.}.
The average value of $\bar{q}(<r)$ decreases with $r$,
reflecting the fact that shocks are stronger on average in the outer regions of clusters,
and the variation of its distribution also decreases with $r$.
The average values, $\left<\bar{q}(<r) \right>\approx 5.5$ at $r = 0$ and 
$\left<\bar{q}(<r) \right>\approx 4.35$ at $r=r_{\rm vir}$,
correspond to the DSA power-law slopes for shocks with $M_s \approx 2$ and 3.5, respectively.
At $r = \rtwo$, the values spread over a narrow range of $\bar{q}(< \rtwo) \approx 4.25 - 4.5$,
corresponding to $M_s \approx 3 - 4$, as shown in the right panel of Figure 10.
The plot indicates that $\bar{q}(< \rtwo)$ does not have any noticeable correlation with
the cluster temperature (nor with the cluster mass and radius although not shown).
These imply that the averaged Mach number of shocks, {\it weighted with CR production},
in our sample clusters is in the range of $\left< M_s \right>_{\rm CR} \approx 3 - 4$,
regardless of the properties of clusters.

\section{Discussion}

The existence of CR protons in galaxy clusters remains to be confirmed directly from observations.
CR protons produce $\gamma$-ray radiation through $p-p$ collisions with background thermal
protons, but so far only upper limits on $\gamma$-ray fluxes from clusters have been set by
Fermi-LAT and VERITAS, as noted in the Introduction \citep{acke10,acke13,arlen12}.
On the other hand, secondary electrons are also produced through $p-p$ collisions, and they
along with $\mu$G-level magnetic fields emit the synchrotron radiation in radio over the
cluster scale.
The observed radio emission is produced typically by electrons with energy of several GeV
corresponding to the Lorentz factor of $\gamma_e \sim 10^4$ \citep{krj2012}.
For this energy range, the secondary electrons {\it at production} have the momentum
distribution similar to the proton spectrum, that is,
$f_e^p(p) \propto p^{-q_e^p}$ with
$q_e^p \approx q$ \citep[e.g.,][]{dermer86}.
For the proton power-law of $\bar{q} \approx 4.25 - 4.5$ estimated in the previous section, 
the secondary electron slope is also $q_e^p \approx 4.25 - 4.5$.
Relativistic electrons suffers radiative coolings, dominantly by synchrotron and inverse
Compton losses.
The cooling time scale of electrons of $\gamma_e \sim 10^4$ is $\tau_{\rm cool} \sim 10^8$
yrs with cluster magnetic fields of a few $\mu$G \citep[e.g.,][]{krj2012}.
The momentum distribution function of the secondary electrons {\it at the steady-state}
governed by the production through $p-p$ collisions and  the cooling processes is given as
$f_e^{ss}(p) \propto p^{-q_e^{ss}}$ with $q_e^{ss} = q_e^p -1$ \citep{de00}.
So for $\bar{q} \approx 4.25 - 4.5$, the secondary electron slope becomes
$q_e^{ss} \approx 5.25 - 5.5$.

Radio halos associated with some galaxy clusters are explained 
as diffuse synchrotron emissions over the cluster scale.
The spectral index of observed radio halos is typically in the range
of $\alpha_R \approx 1 - 1.5$, although in some radio halos it is much steeper \citep{feretti12}.
For $\alpha_R = (q_e-3)/2$, this requires the existence of relativistic electrons 
with the power-slope  $q_e \approx 5 - 6$,
which nicely embraces the slope of steady-state secondary electrons described above.
In the so-called {\it hadronic model}, for instance, the CR electrons emitting synchrotron
radiation are assumed to be secondaries produced through $p-p$ collisions
\citep[e.g.,][]{de00,pfrommer2004}.
Our results indicate that the CR protons accelerated at shocks in the cluster outskirts
may be capable of producing secondary electrons with the right energy spectral slope
($q_e^{ss} \approx 5.25 - 5.5$)
for the spectral index of observed radio halos.
\citet{brunetti2012}, however, argued that at least for the Coma cluster, the hadronic model
that requires the CR proton energy $\ga 3 - 5 \%$ of the thermal energy
may violate the $\gamma$-ray upper limit set by Fermi-LAT observations, 
provided that the magnetic field is not much stronger than that measured/constrained by Faraday RM.
Moreover, according to a more recent Fermi-LAT paper \citep{acke13}, this limit has become even
more stringent, constraining the CR proton energy down to $\la 1 \%$ of the thermal energy
in clusters.
In the so-called {\it re-acceleration model}, on the other hand, the secondary electrons
are further accelerated by turbulence in the ICM, so the CR proton requirement is
alleviated somewhat \citep[e.g.,][]{bsfg01}.
The detailed implications of our results for radio halo are complicated, and addressing them
properly requires studies which are beyond the scope of this paper.

So far, our discussions on DSA at energetic shocks have been focused mostly on CR protons
and secondary electrons resulted from $p-p$ collisions.
As for supernova remnant shocks, primary CR electrons can be accelerated at ICM shocks 
in the same manner as CR protons, although the injection (pre-acceleration) of electrons into DSA
process remains rather uncertain.
We point that the projected surfaces of energetic shocks would have morphologies of
partial shells or elongated arcs (see Figure 4),
so diffuse synchrotron emissions from primary CR electrons accelerated at these shocks could produce
radio structures that resemble radio relics discovered in the cluster outskirts
\citep[e.g.][]{vanweeren2010,krj2012}.
Moreover, the average radial distance of the MESs in Figure 8 is
$\left< r_\mathrm{MES} \right> \sim 0.5\ \rtwo \sim 0.5 - 1.5 \Mpch$, which is
comparable to the average distance of observed radio relics from the cluster center
\citep[e.g.,][]{vanweeren2009}.
So, for instance, radio relics with flat radio spectrum such as the sausage relic in CIZA
J2242.8+5301 ($\alpha_R\approx 0.6$) could be explained by primary electrons accelerated by
energetic shocks (a substantial fraction of which are infall shocks) in the cluster outskirts.

In addition, pre-existing CR electrons in the ICM (previously produced at shocks or
through $p-p$ collisions) can be re-accelerated at energetic shocks \citep{krj2012,pop13}.
To explain the observed properties of radio relics with flat radio spectra, \citet{krj2012},
for instance, proposed a model in which pre-existing CR electrons with the momentum
distribution corresponding to the observed radio spectral index are re-accelerated at weak
shocks with $M_s \la 2 - 3$.
The sausage relic in CIZA J2242.8+5301 then requires pre-existing CR electrons with
$q_e \approx 4.2$.
It is interesting to note that this is close to the slope of the secondary electrons
at production ($q_e^{p} \approx 4.25 - 4.5$) due CR protons accelerated at energetic shocks in
the cluster outskirts.
The electrons with  $\gamma_e \la 10^2$ have the cooling time longer than the age
of the universe.
So we may conjecture a scenario in which the secondary electrons produced with
$\gamma_e \la 10^2$ are boosted to $\gamma_e \ga10^4$ at shocks in the cluster outskirts,
producing radio-relic-like structures.
The acceleration or re-acceleration of CR electrons at shocks in clusters, compared to
those of CR protons, involve additional complications such as injection, pre-existing CR
population, and cooling.

\section{Summary}

The outskirts of galaxy clusters are dynamically active, reflecting disturbances due to mergers
of substructures and continuous infall of gas along filaments.
A manifestation of such activities is the formation of shock waves, which can be observed in
radio and X-ray.
In this paper, we have studied structure formation shocks in the cluster outskirts.
Specifically, in a sample of 228 clusters from numerical simulations of the LSS
formation with uniform grid resolution, we have identified the ICM shocks located in
$r \leq \rtwo\ (\approx 1.3\ \rvir)$
and studied their properties and the CR proton production via DSA there.

Main results are summarized as follows.

1. As previously known \citep[e.g.,][]{ryu2003}, the ICM shocks existing in $r \leq \rtwo$
are mostly weak.
But there are a number of shocks that are strong and energetic enough to produce substantial
amounts of CR protons via DSA.
Shocks with $M_s \approx 3 - 5$ produce the largest amount of CR protons.
Among shocks with $M_s \geq 1.5$, the fractions of shocks (actually shock zones) with
$M_s \geq 3$, $\geq 4$, and $\geq 5$ are $\sim 19 \%$, $\sim 8 \%$, and $\sim 4.5 \%$,
respectively, in our sample.
Shocks with the CR energy flux
$f_\mathrm{CR} \geq 10^{42}\,\mathrm{ergs}\ \mathrm{s}^{-1} (h^{-1}{\rm Mpc})^{-2}$,
were categorized as energetic shocks.
The fraction of energetic shocks (again shock zones) is $\sim 21 \%$ of the identified
shocks.
The shock with the largest $f_{\rm CR}$ in each cluster was designated as the most energetic
shock (MES).

2. Infall shocks, which form by the infall of the WHIM from filaments, were separated from other
shocks by employing the entropy, density, and Mach number criteria.
In $r \le r_{200}$, $\sim 10 \%$ of shocks with $M_s \geq 1.5$ and about a half of shocks with
$M_s \geq 3$ are classified as infall shocks.
Infall shocks are not as common as merger or turbulent shocks.
But with relatively high Mach numbers ($M_s \ga 3$)
and large kinetic energy fluxes, they contribute to a
significant fraction of CR production in clusters.
We found that $\sim 44 \%$ of energetic shocks and $\sim 78 \%$ of the MESs are classified
as infall shocks.
And infall shocks produce $\sim 68 \%$ of CRs in $r \leq \rtwo$, when summed for all
clusters in our sample.

3. Strong energetic shocks, including infall shocks, reside mostly in the cluster outskirts.
Hence, CR protons are expected to be produced mostly in the outskirts and
then advected into the core regions along with flow motions.
The advection time scale is a substantial fraction of the age of the universe.
Consequently, the radial profile of the CR pressure is expected to be broad, dropping off
more slowly than that of the gas pressure, and might be even temporarily inverted peaking
in the outskirts, as shown in previous simulations \citep[e.g.,][]{pfrommer2007,vazza2012}.

4. We have estimated the momentum distribution of the CR protons produced at shocks in
$r \leq \rtwo$.
The volume-integrated momentum spectrum has the plower-law slope of $q_p\approx 4.25 - 4.5$.
So the average Mach number of shocks in our sample clusters, weighted with CR production, 
is in the range of $\left< M_s \right>_{\rm CR} \approx 3 - 4$.
It is greater than the characteristic Mach numbers of merger shocks ($M_s \approx 2-3$).
This confirms that a substantial fraction of CRs are produced by energetic infall shocks
in the cluster outskirts.

5. We suggest that some radio relics with flat radio spectrum 
such as the sausage relic in CIZA J2242.8+5301 ($\alpha_R\approx 0.6$)
could be explained by primary electrons accelerated by energetic infall shocks
with $M_s \ga 3$ induced in the cluster outskirts.

The implications of our results on radio observations of clusters were briefly discussed,
leaving detailed studies for future works.

\begin{acknowledgements}

We thank the anonymous referee for constructive comments.
We also thank G. Brunetti and T. W. Jones for comments on the manuscript.
HK thanks V. Petrosian and KIPAC for their hospitality during the sabbatical leave
at Stanford University, where a part of work was done.
SEH was supported by the National Research Foundation of Korea through grant
2007-0093860.
DR was supported by research fund of Chungnam National University.
HK was supported by Basic Science Research Program through the National Research
Foundation of Korea (NRF) funded by the Ministry of Education, Science and Technology
(2012-001065).
RC was supported in part by grant NASA NNX11AI23G.
Computing resources were in part provided by the NASA High-End Computing (HEC) Program
through the NASA Advanced Supercomputing (NAS) Division at Ames Research Center.

\end{acknowledgements}

\begin{figure}
\hspace{-0.4cm}
\includegraphics[scale=0.8]{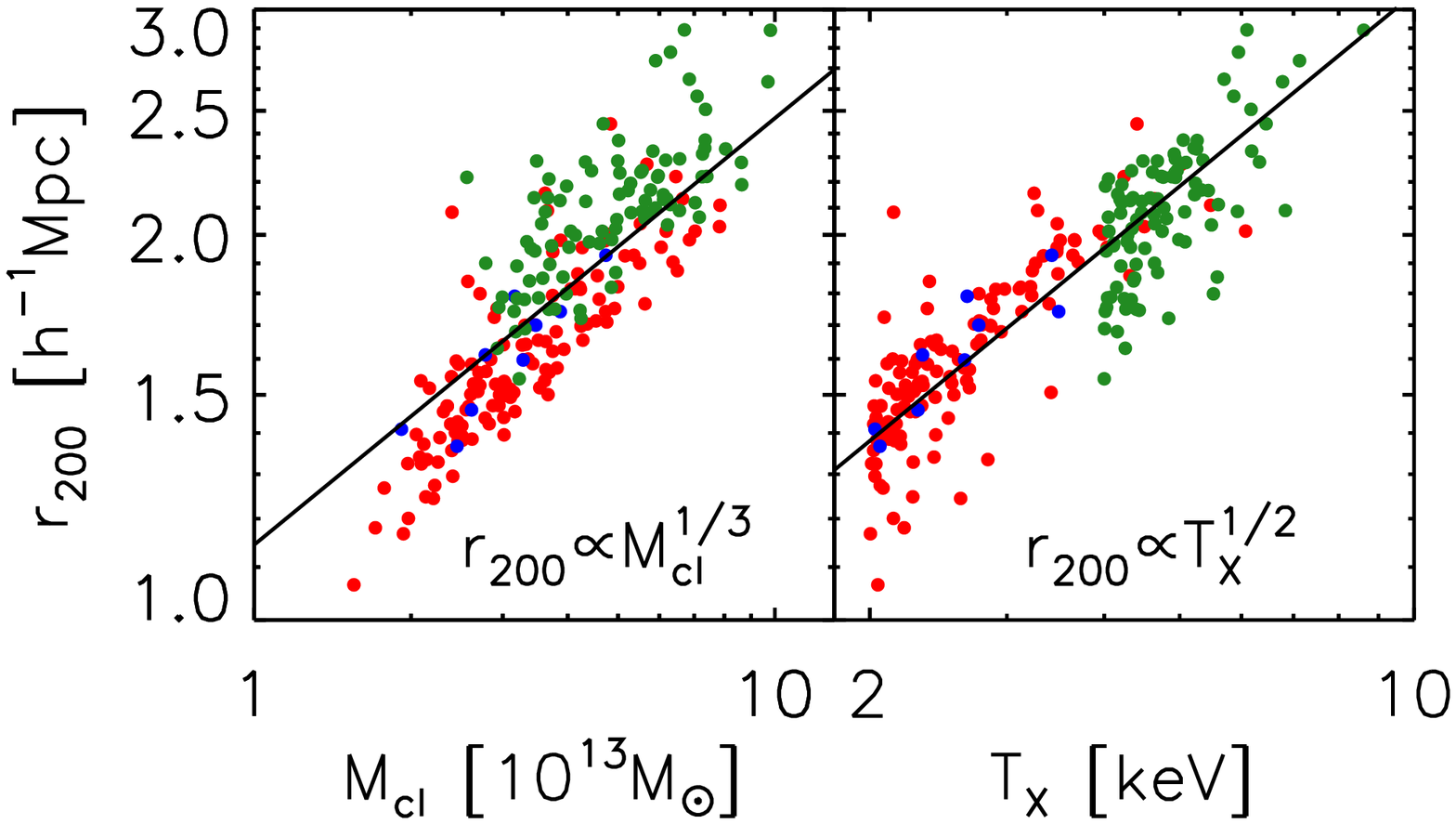}
\caption{Radius as a function gas mass (left) and temperature (right) for clusters
in our sample.
Refer the main text for definitions of $\rtwo$, $M_{\rm cl}$, and $T_{\rm X}$.
Red dots denote 125 clusters with $T_{\rm X} \ge 2$ keV from $100 \Mpch$ box simulations
with $1024^3$ grid zones, green dots denote 94 clusters with $T_{\rm X} \ge 4$ keV from $200 \Mpch$
box simulations with $1024^3$ zones, and blue dots denote 9 clusters with $T_{\rm X} \ge 2$ keV
from $100 \Mpch$ box simulation with $2048^3$ zones, respectively.
Solid lines represent the scaling relations among $\rtwo$, $M_{\rm cl}$, and $T_{\rm X}$, 
expected from virial equilibrium.}
\end{figure}

\begin{figure}
\hspace{1.4cm}
\includegraphics[scale=0.6]{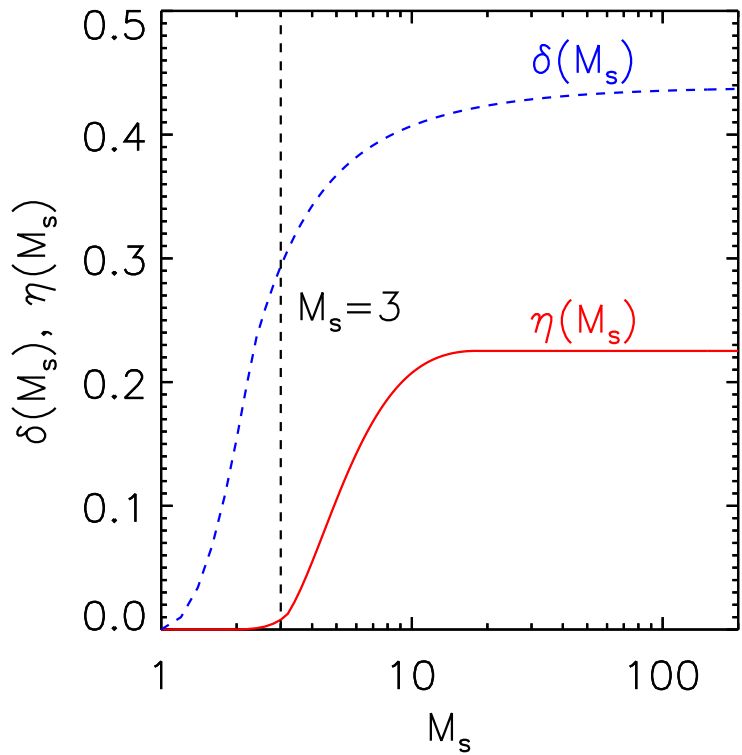}
\caption{Gas thermalization efficiency, $\delta$ (blue dashed curve), and CR-acceleration
efficiency, $\eta$ (red solid curve), employed in this paper, as a function of Mach number.
They were estimated with simulations of nonlinear DSA, where the upstream
$\beta \equiv P_{\rm th}/P_B=100$ was assumed and phenomenological models for MFA and
Alfv\'enic drift in the shock precursor were implemented \citep{kang2013}.
The vertical dashed line marks $M_s = 3$.}
\end{figure}

\begin{figure}
\vspace{-0.1cm}
\hspace{2.8cm}
\includegraphics[scale=0.7]{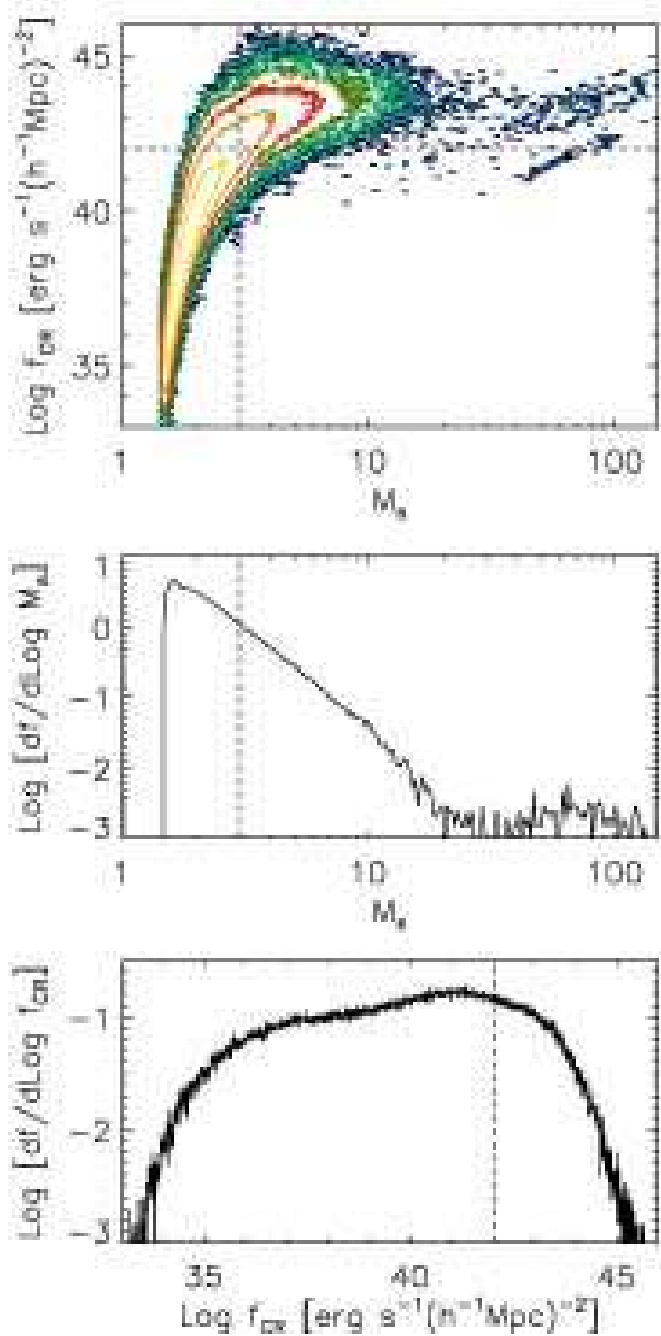}
\vspace{-0.1cm}
\caption{Top: Distribution of shocks (shock zones with assigned $M_s$) found at
$r \leq \rtwo$ in the plane of Mach number and CR energy flux.
The interval of contours is a factor of $\sqrt{10}$.
Middle: Fraction of shocks as a function of Mach number.
Bottom: Fraction of shocks as a function of CR energy flux.
Dashed lines mark $M_s = 3$ and $f_{\rm CR} = 10^{42}$ ergs s$^{-1}$
($h^{-1}$Mpc)$^{-2}$.
The statistics are shown for shocks identified in 134 sample clusters from $100 \Mpch$ box
simulations with $1024^3$ and $2048^3$ grid zones.
}
\end{figure}

\begin{figure}
\vspace{0.cm}
\hspace{-0.5cm}
\includegraphics[scale=0.88]{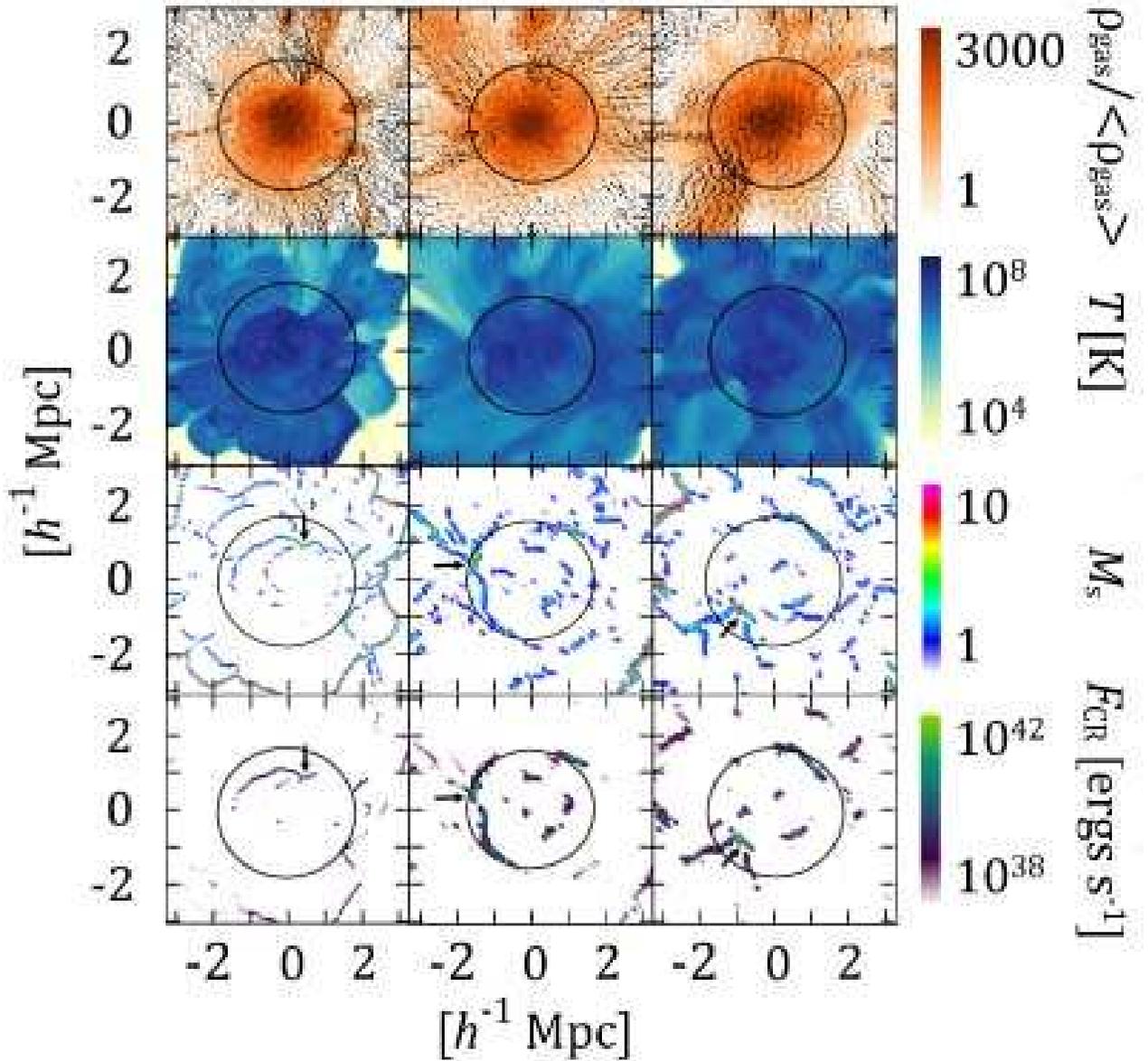}
\vspace{-0.5cm}
\caption{Two-dimensional slice images of three sample clusters showing the gas density with
flow velocity field, gas temperature, shock Mach number, and CR luminosity generated at shocks
(from top to bottom) at present ($z=0$).
The X-ray emission-weighted temperature of clusters is $k T_\mathrm{X} = 2.7$ keV (left),
$2.5$ keV (middle), and $2.4$ keV (right), respectively.
The cluster in the left panels is from $100 \Mpch$ box simulation with $2048^3$ grid zones,
while other two are from $100 \Mpch$ box simulations with $1024^3$ zones.
Circles with $r = \rtwo$ are overlaid in the lower two rows.
Thick arrows in the lower two rows of panels point the most energetic shocks (MESs).
The MESs in the clusters shown here are infall shocks that form in the WHIM
infalling along filaments.}
\end{figure}

\begin{figure}
\vspace{-0.3cm}
\hspace{-0.4cm}
\includegraphics[scale=0.88]{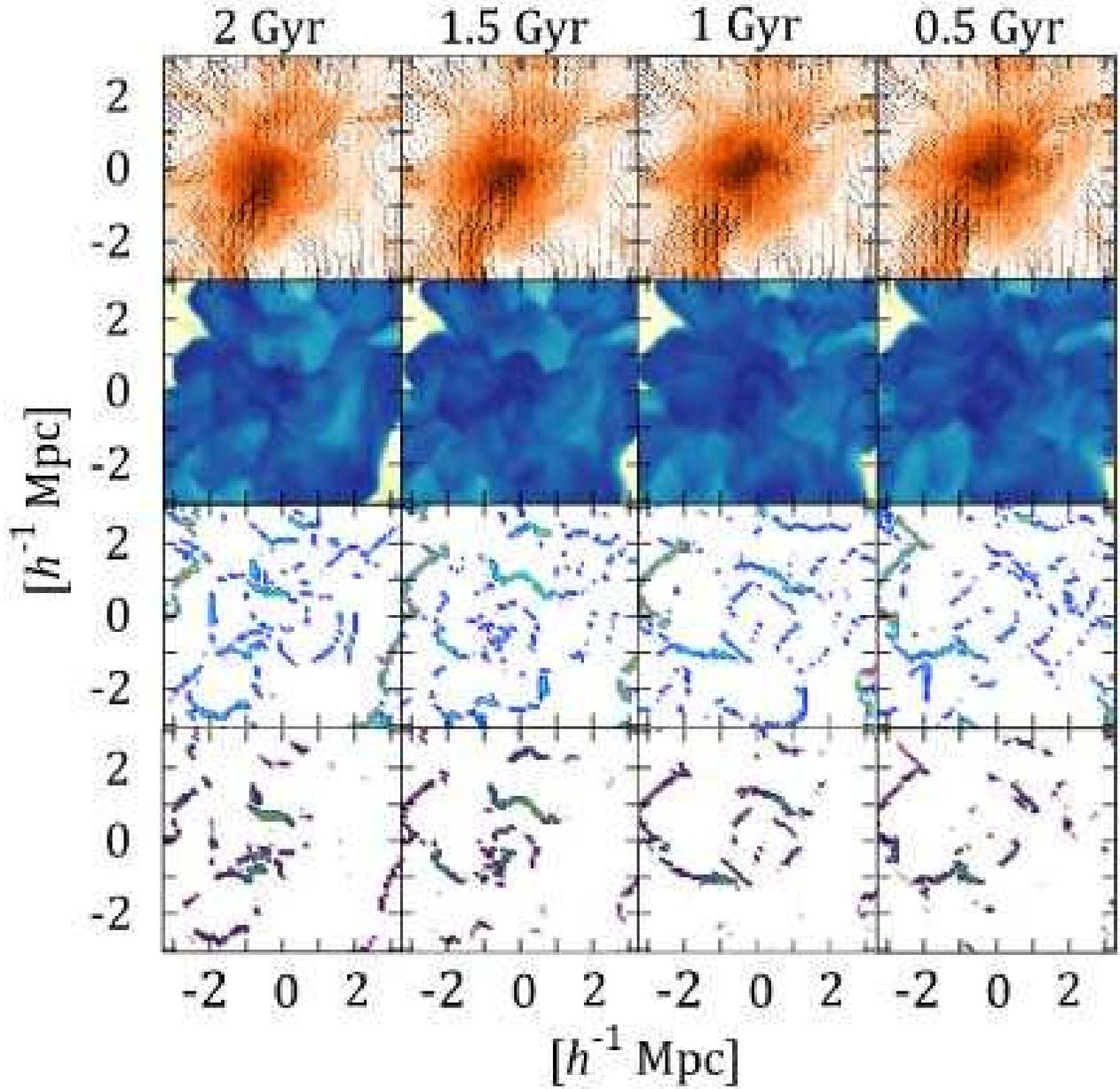}
\vspace{-0.3cm}
\caption{Time evolution of the cluster shown in the right column of Figure 4.
The numbers at the top are the look-back times.
Shown are also the gas density with flow velocity field, gas temperature, shock Mach number,
and CR luminosity generated at shocks (from top to bottom).
The color bars (not shown) are the same as those in Figure 4.}
\end{figure}

\begin{figure}
\hspace{1.cm}
\includegraphics[scale=0.6]{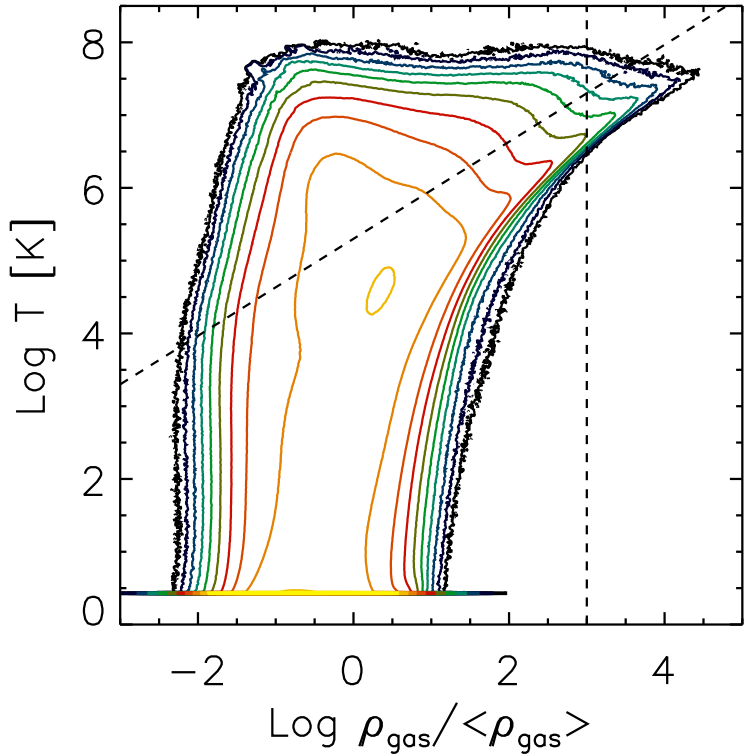}
\caption{Volumetric distribution of the gas in the plane of density and temperature from
$100 \Mpch$ box simulations with $1024^3$ grid zones.
The diagonal and vertical dashed lines mark $\log[T(K)/(\rho/\left<\rho\right>)^{2/3}] = 5.3$
and $\rho/\left<\rho\right> = 10^3$, respectively.
In our classification scheme, infall shocks have the preshock gas 
that is characterized by the following conditions:}
$\log[T(K)/(\rho/\left<\rho\right>)^{2/3}] \le 5.3$ and $\rho/\left<\rho\right> \le 10^3$
(see the main text for details).
\end{figure}

\begin{figure}
\vspace{-0.3cm}
\hspace{0.5cm}
\includegraphics[scale=0.7]{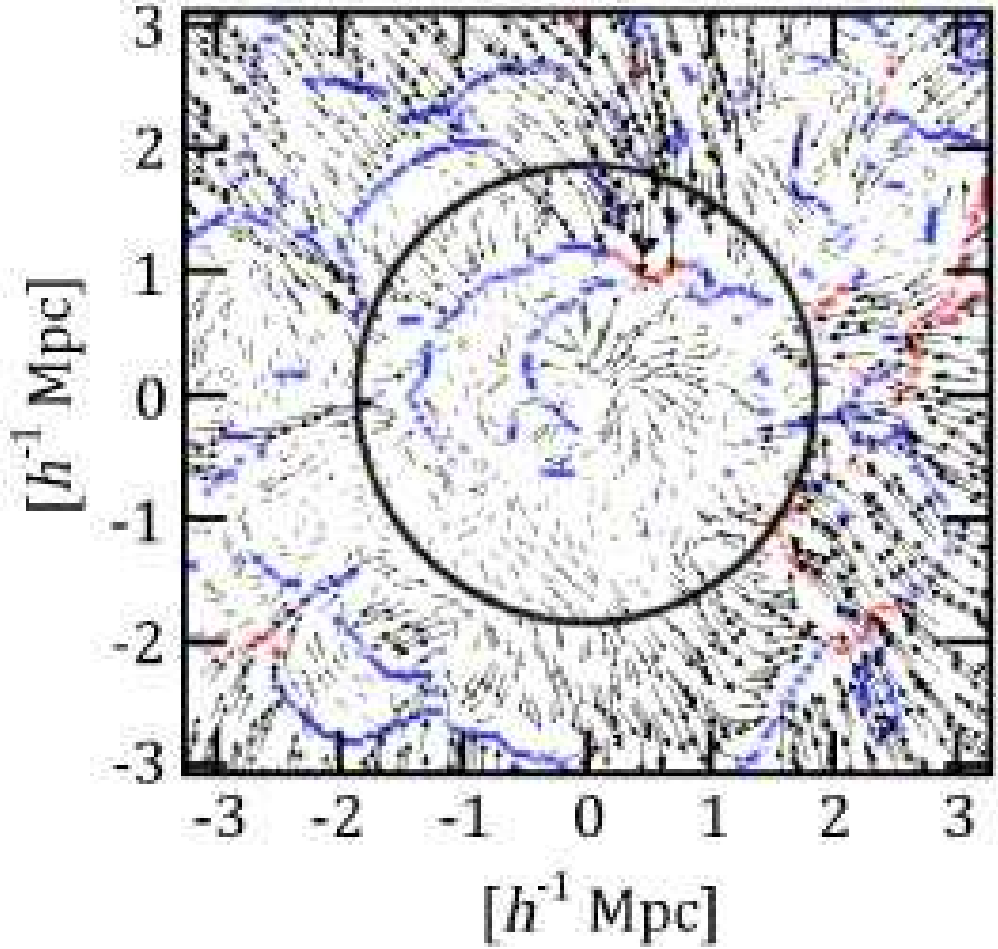}
\vspace{0.cm}
\caption{A slice displaying the positions of infall shocks (red) and not-infall shocks (blue)
according to the criteria in Section 3, along with the flow velocity field, for the cluster shown
in the right column of Figure 4.
The circle marks $r = \rtwo$.}
\end{figure}

\begin{figure}
\hspace{-0.6cm}
\includegraphics[scale=1.05]{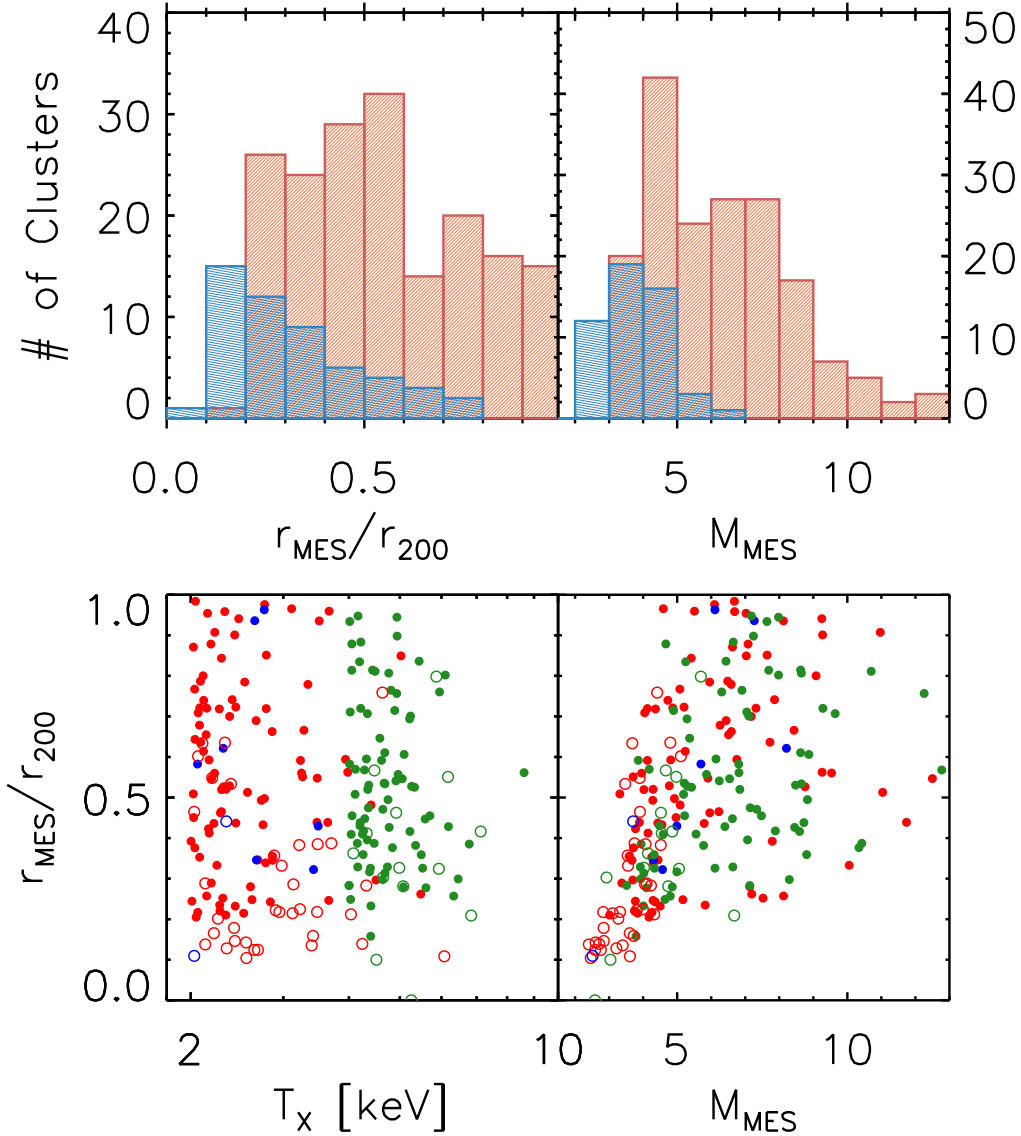}
\vspace{-0.5cm}
\caption{Top: Frequency distributions of radial positions, $r_\mathrm{MES}$ (left), and
Mach numbers, $M_\mathrm{MES}$ (right), of the most energetic shocks (MESs) in our 228
sample clusters.
Red histogram is for 177 clusters where the MESs are infall shocks, and blue histogram
is for 51 clusters where the MESs are not infall shocks.
Bottom: $r_\mathrm{MES}$ as a function of cluster temperature (left) and $M_\mathrm{MES}$
(right).
Red dots denote the clusters from $100 \Mpch$ box simulations with $1024^3$ grid zones, green
dots denote the clusters from $200 \Mpch$ box simulations with $1024^3$ zones, and blue dots
denote the clusters from $100 \Mpch$ box simulation with $2048^3$ zones, respectively.
Filled dots are for 177 clusters in which the MESs are infall shocks, while open dots
are for 51 clusters in which the MESs are not infall shocks.}
\end{figure}

\begin{figure}
\hspace{-0.6cm}
\includegraphics[scale=0.52]{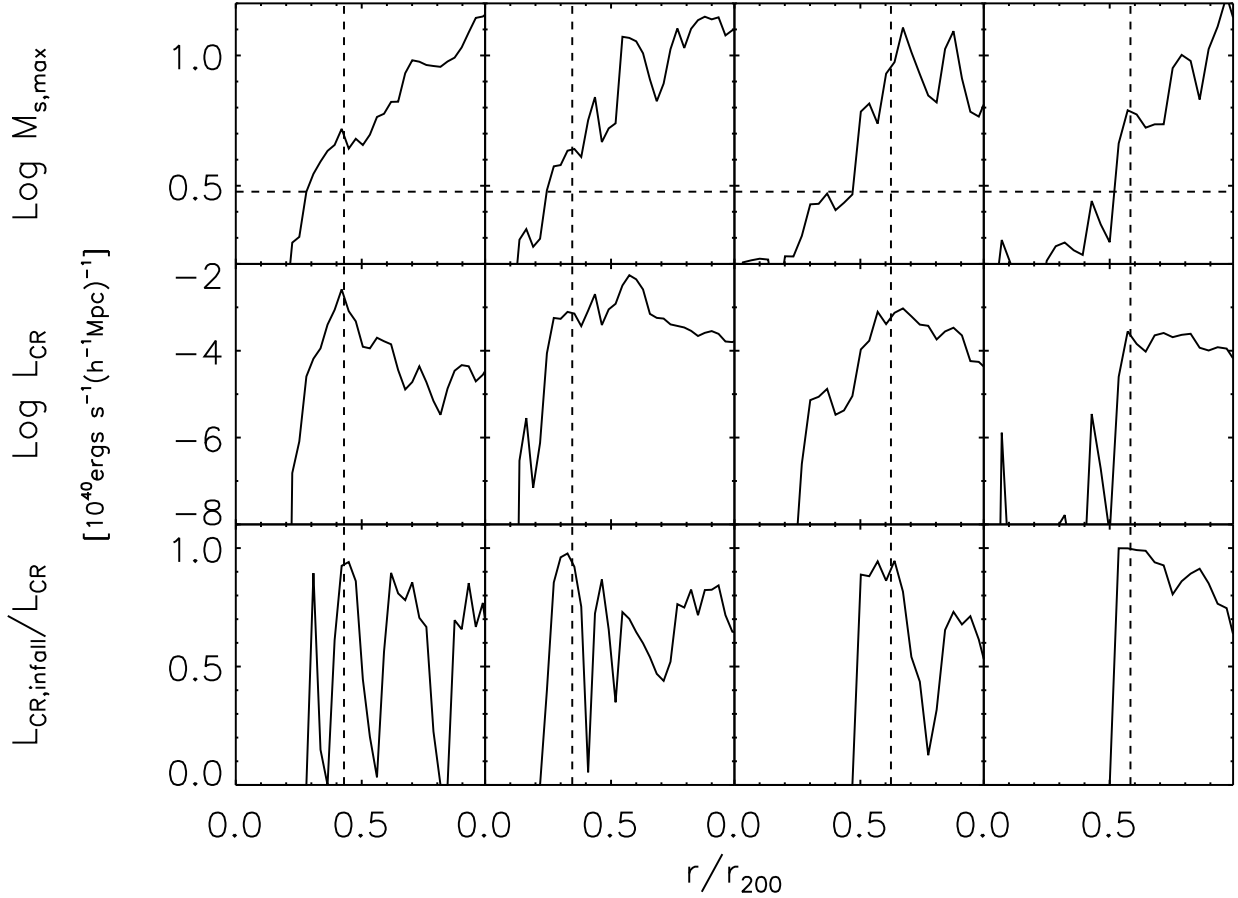}
\caption{Maximum Mach number of shocks, $M_{s,{\rm max}}$ (top), total CR energy
luminosity per unit radius, $L_{\rm CR}$ (middle), and the fraction of CR luminosity due
to infall shocks (bottom), in the radius bin of $[r, r+dr]$, as a function of the
radius in four sample clusters from $100 \Mpch$ box simulations with $1024^3$ grid zones.
The horizontal dashed lines mark $M_{s,{\rm max}} = 3$, while the vertical dashed lines
mark the position of the shells that contain the most energetic shocks (MESs).
The MESs in the clusters shown here are all infall shocks.}
\end{figure}

\begin{figure}
\hspace{-1.4cm}
\includegraphics[scale=0.7]{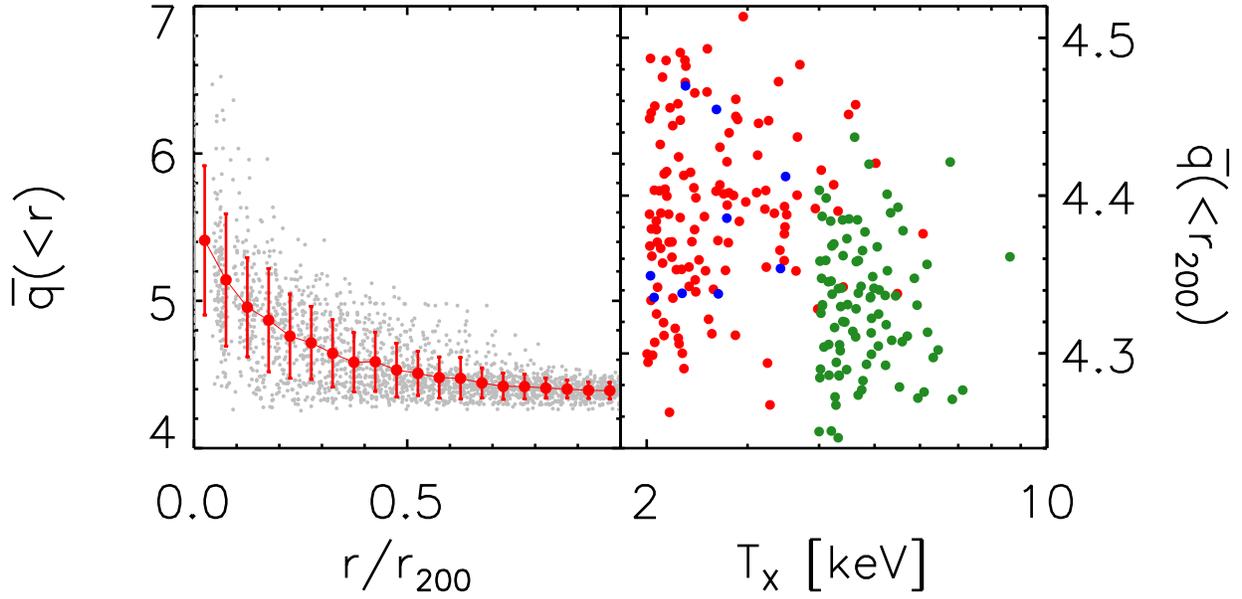}
\vspace{-0.2cm}
\caption{Left: Power-law slope of the momentum distribution of the CR protons produced
at shocks inside the sphere of radius $r$ as a function of $r$.
Grey dots are the data calculated at different radii in some of our sample clusters.
Red filled circles and error bars show their averages and 1 $\sigma$ deviations in given
radius bins, respectively.
Right: Power-law slope of the momentum distribution of the CR protons produced at shocks
inside the sphere of $\rtwo$ as a function of cluster temperature.
Red dots denote the clusters from $100 \Mpch$ box simulations with $1024^3$ grid zones, green
dots denote the clusters from $200 \Mpch$ box simulations with $1024^3$ zones, and blue dots
denote the clusters from $100 \Mpch$ box simulation with $2048^3$ zones, respectively.}
\end{figure}


\begin{thebibliography}{}

\bibitem[Ackermann et al.(2010)]{acke10}
Ackermann, M. et al. 2010, \apjl, 717, L71

\bibitem[Ackermann et al.(2013)]{acke13}
Ackermann, M. et al. 2013, preprint, arXiv:1308.5654

\bibitem[Akamatsu \& Kawahara(2013)]{akamatsu13}
Akamatsu, H., \& Kawahara, H. 2013, \pasj, 65, 16

\bibitem[Akamatsu et al.(2012)]{akamatsu12}
Akamatsu, H., Takizawa, M., Nakazawa, K., Fukazawa, Y., Ishisaki, Y., \& Ohashi, T. 2012,
\pasj, 64, 67

\bibitem[Arlen et al.(2012)]{arlen12}
Arlen, T. et al. 2012, \apj, 757, 123

\bibitem[Bell(1978)]{bell1978}
Bell, A. R. 1978, \mnras, 182, 147 

\bibitem[Bell(2004)]{bell04}
Bell, A. R. 2004, \mnras, 353, 550

\bibitem[Berezinsky et al.(1997)]{bbp97}
Berezinsky, V. S., Blasi, P., \& Ptuskim, V. S. 1997, \apj, 487, 529

\bibitem[Blandford \& Ostriker(1978)]{bo78}
Blandford, R. D., \& Ostriker, J. P. 1978, \apj, 221, L29

\bibitem[Bonafede et al.(2009)]{bonafede2009}
Bonafede, A., Giovannini, G., Feretti, L., Govoni, F., \& Murgia, M. 2009, \aap, 494, 429

\bibitem[Brown \& Rudnick(2011)]{br11}
Brown, S., \& Rudnick, L. 2011, \mnras, 412, 2

\bibitem[Br{\"u}ggen et al.(2012)]{bruggen12}
Br{\"u}ggen M., Bykov A., Ryu D., \& R{\"o}ttgering H., 2012, \ssr, 166, 187

\bibitem[Brunetti et al.(2012)]{brunetti2012}
Brunetti, G., Blasi, P., Reimer, O., Rudnick, L., Bonafede, A., \& Brown, S. 2012,
\mnras, 426, 956 

\bibitem[Brunetti et al.(2001)]{bsfg01} 
Brunetti, G., Setti, G., Feretti, L., \& Giovannini, G. 2001,
\mnras, 320, 365

\bibitem[Bryan \& Norman(1998)]{bn98}
Bryan, G. L., \& Norman, M. L. 1998, \apj, 495, 80

\bibitem[Caprioli(2012)]{caprioli2012}
Caprioli, D. 2012, \jcap, 07, 038

\bibitem[Cen \& Chisari(2011)]{cen2011} 
Cen, R., \& Chisari, N. E. 2011, \apj, 731, 11 

\bibitem[Clarke \& En{\ss}lin(2006)]{clarke2006} 
Clarke, T. E., \& En{\ss}lin, T. A.  2006, \aj, 131, 2900

\bibitem[Dermer(1986)]{dermer86}
Dermer, C. D. 1986, \aap, 157, 223

\bibitem[Dolag \& En{\ss}lin(2000)]{de00}
Dolag, K., \& En{\ss}lin, T. A. 2000, \aap, 362, 151

\bibitem[Drury(1983)]{drury1983}
Drury, L. O'C. 1983, Rept. Prog. Phys., 46, 973

\bibitem[Eke et al.(1998)]{enf98}
Eke, V. R., Navarro, J. F., \& Frenk, C. S. 1998, \apj, 503, 569

\bibitem[En{\ss}lin et al.(2011)]{epms11}
En{\ss}lin, T., Pfrommer, C., Miniati, F., \& Subramanian, K. 2011, \aap, 527, A99

\bibitem[Feretti et al.(2012)]{feretti12}
Feretti, L., Giovannini, G., Govoni, F., \& Murgia, M. 2012, \aapr, 20, 54

\bibitem[Gabici \& Blasi(2003)]{gabici2003}
Gabici, S., \& Blasi, P. 2003, \apj, 583, 695

\bibitem[George et al(2009)]{gfsy09}
George, M. R., Fabian, A. C., Sanders, J. S., Young, A. J., \& Russell, H. R.
2009, \mnras, 395, 657

\bibitem[Hoeft et al.(2008)]{hoeft08} 
Hoeft, M., Br{\"u}ggen, M., Yepes, G., Gottlober, S., \& Schwope, A. 2008, \mnras, 391, 1511

\bibitem[Kang(2012)]{kang2012}
Kang, H. 2012, J. Korean Astron. Soc., 45, 127

\bibitem[Kang et al.(1994)]{kang1994} 
Kang, H., Cen R., Ostriker, J. P., \& Ryu, D. 1994, \apj, 428, 1

\bibitem[Kang \& Jones(2007)]{kj07}
Kang, H., \& Jones, T. W. 2007, Astropart. Phys., 28, 232

\bibitem[Kang \& Ryu(2010)]{kang2010}
Kang, H., \& Ryu, D., 2010, \apj, 721, 886

\bibitem[Kang \& Ryu(2011)]{kang2011}
Kang, H., \& Ryu, D., 2011, \apj, 734, 18

\bibitem[Kang \& Ryu(2013)]{kang2013}
Kang, H., \& Ryu, D., 2013, \apj, 764, 95

\bibitem[Kang et al.(2007)]{kang2007} 
Kang, H., Ryu, D., Cen, R., \& Ostriker, J. P. 2007, \apj, 669, 729

\bibitem[Kang et al.(2012)]{krj2012} 
Kang, H., Ryu, D., \& Jones, T. W. 2012, \apj, 756, 97 

\bibitem[Komatsu et al.(2011)]{komatsu2011} 
Komatsu, E., et al. 2011, \apjs, 192, 18 

\bibitem[Lau et al.(2009)]{lau2009}
Lau E. T., Kravtsov A. V., \& Nagai D., 2009, \apj, 705, 1129

\bibitem[Li et al.(2008)]{llc08}
Li, S., Li, H., \& Cen, R. 2008, \apjs, 174, 1

\bibitem[Lucek \& Bell(2000)]{lucek00}
Lucek, S. G., \& Bell, A. R. 2000, \mnras, 314, 65

\bibitem[Markevitch et al.(1998)]{mfsv98}
Markevitch, M., Forman, W. R., Sarazin, C. L., \& Vikhlinin, A. 1998, \apj, 503, 77

\bibitem[Markevitch et al.(2002)]{markevitch02}
Markevitch, M., Gonzalez, A. H., David, L., Vikhlinin, A., Murray, S., Forman, W., Jones, C.,
\& Tucker, W. 2002, \apj, 567, 27

\bibitem[Markevitch \& Vikhlinin(2007)]{markevitch07}
Markevitch, M., \& Vikhlinin, A. 2007, \physrep, 443, 1

\bibitem[Miniati et al.(2001)]{miniati2001}
Miniati, F., Ryu, D., Kang, H., \& Jones, T. W. 2001, \apj, 559, 59

\bibitem[Minati et al.(2000)]{miniati2000}
Miniati, F., Ryu, D., Kang, H., Jones, T. W., Cen, R., \& Ostriker, J. 2000, \apj, 542, 608

\bibitem[Nakamura \& Suto(1997)]{ns97}
Nakamura, T. T., \& Suto, Y. 1997, Prog. of Theor. Phys., 97, 49

\bibitem[Nuza et al.(2012)]{nuza2012}
Nuza, S. E., Hoeft, M., van Weeren, R. J., Gottl{\"o}ber, S., \& Yepes, G. 2012,
\mnras, 420, 2006

\bibitem[Ogrean \& Br{\"u}ggen(2013)]{ob13}
Ogrean, G. A., \& Br{\"u}ggen, M. 2013, \mnras, 433, 1701

\bibitem[Ogrean et al.(2013)]{ogrean13}
Ogrean, G. A., Br{\"u}ggen, M., van Weeren, R. J., R{\"o}ttgering, H., Croston, J. H.,
\& Hoeft, M. 2013, \mnras, 433, 812

\bibitem[Peebles(1980)]{peeb80}
Peebles P. J. E. 1980, The Large-Scale Structure of the Universe
(Princeton: Princeton Univ. Press) 

\bibitem[Pfrommer \& En{\ss}lin(2004)]{pfrommer2004}
Pfrommer, C., \& En{\ss}lin, T. A., 2004, \mnras, 352, 76

\bibitem[Pfrommer et al.(2007)]{pfrommer2007}
Pfrommer, C., En{\ss}lin, T. A., Springel, V., Jubelgas, M., \& Dolag, K. 2007, \mnras, 378, 385

\bibitem[Pfrommer \& Jones(2011)]{pj11}
Pfrommer, C., \& Jones, T. W. 2011, \apj, 730, 22

\bibitem[Pfrommer et al.(2006)]{psej06}
Pfrommer, C., Springel, V., En{\ss}lin, T. A., \& Jubelgas, M. 2006, \mnras, 367, 113

\bibitem[Pinzke et al.(2013)]{pop13}
Pinzke, A., Oh, S. P., \& Pfrommer, C. 2013 \mnras, 435, 1061 

\bibitem[Russell et al.(2010)]{russell10} 
Russell, H. R., Sanders, J. S., Fabian, A. C., Baum, S. A., Donahue, M., Edge, A. C., 
McNamara, B. R., \& O'Dea, C. P. 2010, \mnras, 406, 1721

\bibitem[Ryu et al.(2008)]{ryu2008}
Ryu, D., Kang, H., Cho, J., \&  Das, S. 2008, Science, 320, 909

\bibitem[Ryu et al.(2003)]{ryu2003} 
Ryu, D., Kang, H., Hallman, E., \& Jones, T. W. 2003, \apj, 593, 599

\bibitem[Ryu et al.(1993)]{ryu1993}
Ryu, D., Ostriker, J. P., Kang, H., \& Cen, R. 1993, \apj, 414, 1

\bibitem[Ryu et al.(2012)]{ryu2012}
Ryu, D., Schleicher, D. R. G., Treumann, R. A., Tsagas, C. G., \& Widrow, L. M. 2012,
\ssr, 166, 1

\bibitem[Schure et al.(2012)]{schure2012}
Schure, K. M., Bell, A. R, Drury, L. O'C., \&. Bykov, A. M. 2012, \ssr, 173, 491

\bibitem[Simionescu et al.(2013)]{simionescu2013}
Simionescu, A., Werner, N., Urban, O., Allen, S. W., Fabian, A. C., Mantz, A., Matsushita, K.,
Nulsen, P. E. J., Sanders, J. S., Sasaki, T., Sato, T., Takei, Y., \& Walker, S. A. 2013,
\apj, 775, 4

\bibitem[Skillman et al.(2011)]{skillman2011} 
Skillman, S. W., Hallman, E. J., O'Shea, B. W., Burns, J. O., Smith, B. D., \& Turk, M. J.
2011, \apj, 735, 96

\bibitem[Skillman et al.(2008)]{skillman2008} 
Skillman, S. W., O'Shea, B. W., Hallman, E. J., Burns, J. O., \& Norman, M. L. 2008,
\apj, 689, 1063

\bibitem[Skillman et al.(2013)]{skillman2013} 
Skillman, S. W., Xu, H., Hallman, E. J., O'Shea, B. W., Burns, J. O., Li, H., Collins, D. C.,
\& Norman, M. L. 2013, \apj, 765, 21

\bibitem[van Weeren et al.(2009)]{vanweeren2009} 
van Weeren, R. J., R{\"o}ttgering, H. J. A., Br{\"u}ggen, M., \& Cohen, A. 2009,
\aap, 508, 75 

\bibitem[van Weeren et al.(2010)]{vanweeren2010} 
van Weeren, R. J., R{\"o}ttgering, H. J. A., Br{\"u}ggen, M., \& Hoeft, M. 2010,
Science, 330, 347 

\bibitem[van Weeren et al.(2012)]{vanweeren2012}
van Weeren, R. J., R{\"o}ttgering, H. J. A., Intema, H. T., Rudnick, L., Br{\"u}ggen, M.,
Hoeft, M., \& Oonk, J. B. R. 2012, \aap, 546, A124

\bibitem[Vazza et al.(2012)]{vazza2012}
Vazza, F., Br{\"u}ggen, M., Gheller, C., \& Brunetti, G. 2012 \mnras, 421, 3375 

\bibitem[Vazza et al.(2009a)]{vazza2009a}
Vazza, F., Brunetti, G, \& Gheller, C. 2009a, \mnras, 395, 1333

\bibitem[Vazza et al.(2009b)]{vazza2009b}
Vazza, F., Brunetti, G., Kritsuk, A., Wagner, R., Gheller, C., \& Norman, M.
2009b, \aap, 504, 33

\bibitem[Vazza et al.(2011)]{vazza2011}
Vazza, F., Dolag, K., Ryu, D., Brunetti, G., Gheller, C., Kang, K., \& Pfrommer, C.
2011, \mnras, 418, 960

\bibitem[Vikhlinin et al.(2006)]{vkfj06}
Vikhlinin, A., Kravtsov, A., Forman, W., Jones, C., Markevitch, M., Murray, S. S.,
\& Van Speybroeck, L. 2006, \apj, 640, 691

\bibitem[Voit et al.(2005)]{vkb05}
Voit, G. M., Kay, S. T., \& Bryan, G. L. 2005, \mnras, 364, 909

\bibitem[Walker et al.(2012)]{walker2012}
Walker, S. A., Fabian, A. C., Sanders, J. S., \& George, M. R. 2012, \mnras, 427, L45

\end{thebibliography}
\end{document}